\documentclass{article}

\usepackage[usenames,dvipsnames,svgnames,table]{xcolor}
\definecolor{pdfbgcolor}{RGB}{180,180,180}

\usepackage{hyperref}

\usepackage{etoolbox}
\usepackage{environ}
\usepackage{xspace}
\newtoggle{isthesis}
\togglefalse{isthesis}
\NewEnviron{thesisonly}{%
  \iftoggle{isthesis}
    {\BODY\xspace}
    {}%
  }
\NewEnviron{thesisnot}{%
  \iftoggle{isthesis}
    {}
    {\BODY\xspace}%
  }
  
\NewEnviron{fullver}{%
  \BODY\xspace%
}
\NewEnviron{shortver}{}

\usepackage{tabto}
\usepackage{comment}
\excludecomment{ignore}
\usepackage{amsfonts}
\usepackage{amssymb}
\usepackage{graphicx}
\usepackage{setspace}
\usepackage{smartref} 
\usepackage{fullpage}
\usepackage{listings}
\usepackage{enumerate}
\usepackage{rotating}
\usepackage[normalem]{ulem}
\usepackage{paralist}
\usepackage{wrapfig}
\usepackage{paralist}
\usepackage{tabularx}
\usepackage[shortlabels]{enumitem}
\usepackage{framed}
\lstdefinestyle{nonumbers}{numbers=none}

\newcommand{\ourcomments}[2]{}

\newcommand{\trevor}[1]{\ourcomments{#1}{Trevor}}

\usepackage{tikz}

\usepackage[T1]{fontenc}
\usepackage[utf8]{inputenc}
\usepackage{mathptmx}
\usepackage{pifont}

\newcommand{\cmark}{\checkmark}

\lstdefinestyle{customc}{
  belowcaptionskip=1\baselineskip,
  breaklines=true,
  frame=L,
  xleftmargin=\parindent,
  language=C++,
  showstringspaces=false,
  basicstyle=\scriptsize\ttfamily,
  keywordstyle=\bfseries\color{green!40!black},
  commentstyle=\itshape\color{purple!40!black},
  identifierstyle=\color{blue},
  stringstyle=\color{orange},
  numbers=left
}

\newcommand{\record}{record}

\newcommand\leaveq{\textit{leaveQstate}}
\newcommand\enterq{\textit{enterQstate}}
\newcommand\retire{\textit{retire}}
\newcommand\isq{\textit{isQuiescent}}

\begin{document}

%
%
%
%

\title{Reclaiming Memory for Lock-Free Data Structures: \\ There has to be a Better Way}

\author{Trevor Brown \\ Institute of Science and Technology, Austria \\ me@tbrown.pro}
\date{}




\maketitle

\begin{abstract}
Memory reclamation for sequential or lock-based data structures is typically easy.
However, memory reclamation for lock-free data structures is a significant challenge.
Automatic techniques such as garbage collection are inefficient or use locks, and non-automatic techniques either have high overhead, or do not work for many reasonably simple data structures.
For example, subtle problems can arise when hazard pointers, one of the most common non-automatic techniques, are applied to many natural lock-free data structures.
Epoch based reclamation (EBR), which is by far the most efficient non-automatic technique, allows the number of unreclaimed objects to grow without bound, because one slow or crashed process can prevent all other processes from reclaiming memory.

We develop a more efficient, distributed variant of EBR that solves this problem.
It is based on signaling, which is provided by many operating systems, such as Linux and UNIX.
Our new scheme takes O(1) amortized steps per high-level operation on the lock-free data structure and O(1) steps in the worst case each time an object is removed from the data structure.
At any point, $O(mn^2)$ objects are waiting to be freed, where $n$ is the number of processes and $m$ is a small constant for most data structures.
Experiments show that our scheme has very low overhead: on average 10\%, and at worst 28\%, for a balanced binary search tree over many thread counts, operation mixes and contention levels.
Our scheme also outperforms a highly efficient implementation of hazard pointers by an average of 75\%.

Typically, memory reclamation code is tightly woven into lock-free data structure code.
To improve modularity and facilitate the comparison of different memory reclamation schemes, 
we also introduce a highly flexible abstraction.
It allows a programmer to easily interchange schemes for reclamation, object pooling, allocation and deallocation with virtually no overhead, by changing a single line of code.
\end{abstract}



\begin{thesisnot}
\section{Introduction} \label{sec-intro}

In concurrent data structures that use locks, it is typically straightforward to free memory to the operating system after a node is removed from the data structure.
For example, consider a singly-linked list implementation of the set abstract data type using hand-over-hand locking.
Hand-over-hand locking allows a process to lock a node (other than the head node) only if it holds a lock on the node's predecessor.
To traverse from a locked node $u$ to its successor $v$, the process first locks $v$, then it unlocks $u$.
To delete a node $u$, a process first locks the head node, then performs hand-over-hand locking until it reaches $u$'s predecessor and locks it.
The process then locks $u$ (to ensure no other process holds a lock on $u$), removes it from the list, frees it to the operating system, and finally unlocks $u$'s predecessor.
It is easy to argue that no other process has a pointer to $u$ when $u$ is freed.

In contrast, memory reclamation is one of the most challenging aspects of lock-free data structure design.
Lock-free algorithms (also called non-blocking algorithms) guarantee that as long as some process continues to take steps, eventually some process will complete an operation.
The main difficulty in performing memory reclamation for a lock-free data structure is that a process can be sleeping while holding a pointer to an object that is about to be freed.
Thus, carelessly freeing an object can cause a sleeping process to access freed memory when it wakes up, crashing the program or producing subtle errors.
Since nodes are not locked, processes must coordinate to let each other know which nodes are safe to reclaim, and which might still be accessed.
(This is also a problem for data structures with lock-based updates and lock-free queries.
The solutions presented hererin apply equally well to lock-based data structures with lock-free queries.)
%
\end{thesisnot}

\begin{thesisonly}%
In this chapter, we study the problem of performing \textit{safe memory reclamation} for lock-free data structures.
\end{thesisonly}
\begin{wrapfigure}{r}{0.5\textwidth}
\includegraphics[width=\linewidth]{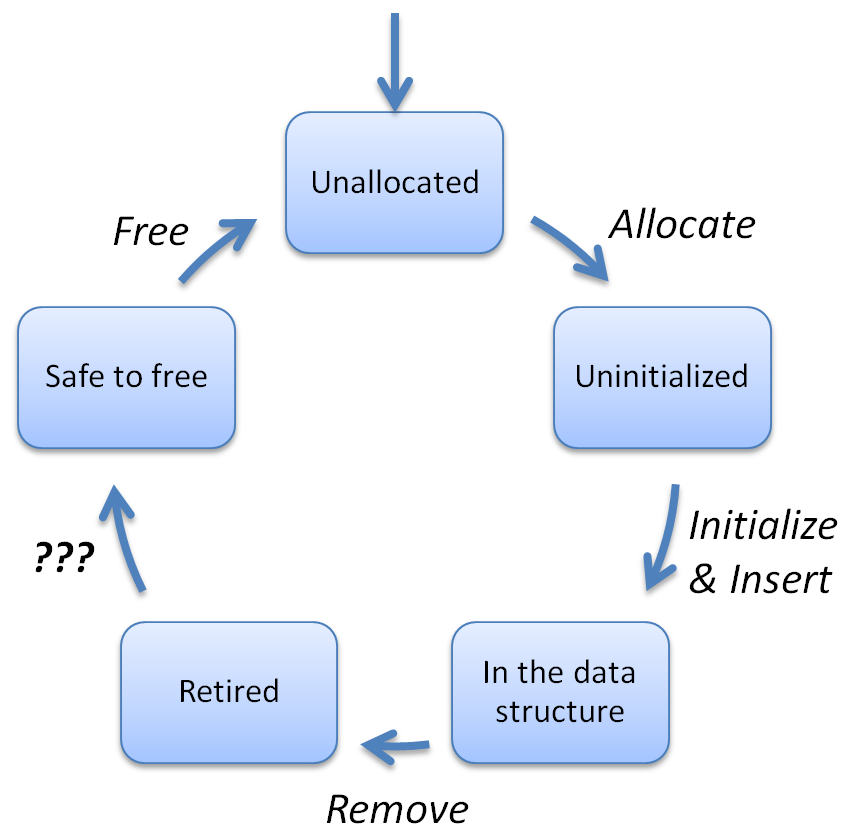}
\caption{The lifecycle of a record.}
\label{fig-lifecycle-of-record}
\end{wrapfigure}
%
Consider a linked data structure that contains \textit{records}, which point to one another.
Figure~\ref{fig-lifecycle-of-record} shows the lifecycle of a record.
Initially all records are \textit{unallocated}.
A process can \textit{allocate} a record, after which we say the record is \textit{uninitialized}.
The process then initializes and inserts the record into the data structure.
Eventually, a process may remove the record from the data structure, after which we say the record is \textit{retired}.
The goal of safe memory reclamation is to determine when it is safe to \textit{free} a retired record, returning it to an unallocated state.
Once it is safe to free a record, one can either free it, or immediately reuse it.
In either case, we say that the record has been \textit{reclaimed}.

It is conceptually useful to divide the work of reclaiming an object into two parts: determining when a record is retired, and determining when it is safe to free (because no process can reach it by following pointers).
We call a memory reclamation scheme \textit{automatic} if it performs all of the work of reclaiming a record, and \textit{non-automatic} if it requires a data structure operation to invoke a procedure whenever it removes a record from the data structure.
Non-automatic memory reclamation schemes may also require a data structure operation to invoke procedures when other events occur, for instance, when the operation begins, or when it accesses a new record.
A lock-free algorithm can invoke these procedures only if they are lock-free or wait-free.
(Otherwise, they will break the lock-free progress property.)

To specify progress for a memory reclamation scheme, it is also necessary to provide some guarantee that records will eventually be reclaimed.
Ideally, one would guarantee that \textit{every} retired record is eventually reclaimed.
However, it is impossible to guarantee such a strong property when processes can crash.
Since it is not safe to free a record while a process has a pointer to it, and one cannot distinguish between a crashed process and a very slow one, a crashed process can prevent some records from ever being reclaimed.
%
We call a memory reclamation scheme \textit{fault-tolerant} if crashed processes can only prevent a bounded number of records from being reclaimed.

Automatic memory reclamation offers a simple programming environment, but can be inefficient, and is typically not fault-tolerant.
Non-automatic memory reclamation schemes are particularly useful when developing data structures for software libraries, since optimizations can pay large dividends when code is reused.
Additionally, non-automatic techniques can be used to build data structures that serve as building blocks to construct new automated techniques.

Garbage collectors comprise the largest class of automatic memory reclamation schemes. 
The literature on garbage collection is vast, and lies outside the scope of this work.
Garbage collection schemes are surveyed in \cite{Jones1996, Schoeberl2010}.
Reference counting is another memory reclamation scheme that can be automatic.
Limited forms of reference counting are also used to construct non-automatic memory reclamation schemes.

Non-automatic techniques can broadly be grouped into five categories: \textit{unsafe reclamation}, \textit{reference counting}, \textit{hazard pointers}, \textit{epoch based reclamation} and \textit{transactional memory assisted reclamation}.
Unsafe reclamation algorithms do not implement safe memory reclamation.
Instead, they immediately reclaim records without waiting until they can safely be freed, which can cause numerous problems.
For example, suppose a process $p$ reads a field $f$ of a record $r$ and sees A, then performs a
\begin{thesisnot}
compare-and-swap (CAS)
\end{thesisnot}
\begin{thesisonly}
CAS
\end{thesisonly}
instruction to change $f$ from A to B. 
If $r$ is reclaimed before it can safely be freed, then $p$ can still have a pointer to $r$ \textit{after} $r$ has been reclaimed.
Thus, $p$'s CAS can be performed after $r$ has been reclaimed.
And, if $f$ happens to contain A when $p$'s CAS is performed, then the CAS will erroneously succeed, effectively changing a different record than the one $p$ intended to change.
Unsafe reclamation algorithms must ensure that such problems do not occur. 
Algorithms in the other categories implement safe memory reclamation, and cannot experience such problems.
As will be discussed in Section~\ref{sec-debra-related}, 
existing techniques either do not work for, or are inefficient for, many natural lock-free data structures. 

In 
epoch based reclamation (EBR), the execution is divided into epochs.
Each record removed from the data structure in epoch $e$ is placed into a shared \textit{limbo bag} for epoch $e$.
Limbo bags are maintained for the last three epochs. 
Each time a process starts an operation, it reads and announces the current epoch, and checks the announcements of other processes.
If all processes have announced the current epoch, then a new epoch begins, and the contents of the oldest limbo bag can be reclaimed.
If a process sleeps or crashes during an operation, then no memory can be reclaimed, so EBR is not fault tolerant.

The first contribution of this work is DEBRA, a distributed variant of EBR with numerous advantages over classical EBR.
DEBRA supports \textit{partial fault tolerance} by allowing reclamation to continue after a process crashes, as long as that process was not performing an operation on the data structure.
DEBRA also significantly improves the performance of EBR by: amortizing the cost of checking processes' announced epochs over many operations, eliminating the shared limbo bags in favour of private limbo bags for each process, and optimizing for good cache performance (even on NUMA systems).
DEBRA performs O(1) steps at the beginning and end of each data structure operation and O(1) steps each time an record is removed from the data structure.

%

Our main contribution is DEBRA+, the first fault tolerant epoch based reclamation scheme.
The primary challenge was to find a way to allow the epoch to be advanced without waiting for slow processes that have not announced the current epoch after a long time, and which may have crashed.
In order to advance the epoch, we must ensure that such a process will not access any retired record if it takes another step.
The technique we introduce for adding fault tolerance to DEBRA uses signals, an interprocess communication mechanism supported by many operating systems (e.g., Linux and UNIX).
With DEBRA+, at most $O(mn^2)$ records are in the limbo bags, waiting to be freed, where $n$ is the number of processes and $m$ is the largest number of records removed from the data structure by one operation. 

A major problem arises when auditioning non-automatic techniques to see which performs best 
for a given lock-free data structure. 
The set of operations exposed to the programmer by non-automatic techniques varies widely.
This is undesirable for several reasons.
First, from a theoretical standpoint, there is no reason that a data
structure should have to be aware of how its records are allocated and reclaimed.
Second, it is difficult to interchange memory reclamation schemes to determine
which performs best in a given situation.
Presently, doing this entails writing many
different versions of the data structure, each version tailored to a specific memory
reclamation scheme.
Third, as new advancements in memory reclamation appear, existing data structures have to be reimplemented (and their correctness painstakingly re-examined) before they can reap the benefits.
Updating lock-free data structures in this way is non-trivial.
Fourth, moving code to a machine with a different memory consistency model can require changes to the memory reclamation scheme, which currently requires changing the code for the lock-free data structure as well.
Fifth, if several instances of a data structure are used for very different purposes (e.g., many small trees with strict memory footprint requirements and one large tree with no such requirement), then it may be appropriate to use different memory reclamation schemes for the different instances.
Currently, 
this requires writing and maintaining code for different versions of the data structure.

These issues were all considered in the context of sequential data structures when
the C++ standard template libraries were implemented.
Their solution was to introduce an \textit{Allocator} abstraction, which allows a data structure to perform \textit{allocate} and \textit{deallocate} operations. 
With this abstraction, the memory allocation scheme for a data structure
can easily be changed without modifying 
the data structure code at all.
Unfortunately, the Allocator abstraction cannot be applied directly to lock-free programming, since its operations do not align well with the operations of lock-free memory reclamation schemes.
(For example, it requires the data structure to know when it is safe to free a record.)
In Section~\ref{sec-abstraction}, we present our third contribution, the first \textit{lock-free Allocator} abstraction.
The main challenge in developing such an abstraction is to find the right set of operations to expose to the data structure implementation, so that the result is simultaneously highly efficient, versatile and easy to program.

Experiments on C++ implementations of DEBRA and DEBRA+ show that overhead is very low.
Compared with performing \textit{no reclamation} at all, DEBRA is on average 4\% slower, and at worst 21\% slower, over a wide variety of thread counts and workloads.
In some experiments, DEBRA actually \textit{improves} performance by as much as 20\%.
Although it seems impossible to achieve better performance when computation is spent reclaiming records, DEBRA reduces the memory footprint of the data structure, which improves memory locality and cache performance.
Adding fault tolerance to DEBRA adds 2.5\% overhead, on average.
Section~\ref{sec-debra-exp} presents extensive experiments comparing DEBRA and DEBRA+ with other leading memory reclamation schemes.
For example, DEBRA+ also outperforms a highly efficient implementation of hazard pointers by an average of 75\%. 


\begin{thesisnot}
\section{Model} \label{sec-prelim}

We consider an asynchronous shared memory system with $n$ processes.
Each process has local memory that is not accessible by any other process, and there is a shared memory accessible by all processes.
Memory is divided into primitive objects, which have atomic operations that are provided directly by the hardware.
Examples include read/write registers, compare-and-swap (CAS) objects, and double-wide compare-and-swap (DWCAS) objects.
A \textit{\record} is a collection of primitive objects, which we refer to as \textit{fields}.
A data structure consists of a fixed set of \textit{entry points}, which are pointers to \record s, and the \record s that are reachable by following one or more pointers from an entry point.
The system has an memory \textit{allocator} that provides operations to \textit{allocate} and \textit{free} \record s.
Initially, all \record s in shared memory are \textit{unallocated}.
Accessing an unallocated record will cause program failure.
Allocating a \record\ provides the process that requested it with a pointer to it, and makes it accessible by any process that has a pointer to it.
A \record\ can also be \textit{freed}, which returns it to the \textit{unallocated} state.

\paragraph{Cache coherence}


We consider a standard modified-exclusive-shared-invalid (MESI) cache coherence protocol.
Whenever a process $p$ attempts to access a cache line, it checks each of its caches one-by-one, starting from the highest level, to see whether the cache line already resides in any of its caches.
At each level of the cache hierarchy where $p$ fails to find the desired cache line, we say that it experiences a \textit{cache miss}.
The last cache level before main memory is called the \textit{last-level cache}, and a cache miss at that level is called a \textit{last-level cache miss}.
If $p$ experiences a last-level cache miss, it must retrieve a copy of the cache line from main memory.

When reading from memory, a process loads a cache line into its cache in \textit{shared} mode.
Many processes can simultaneously have the same cache line in their private caches, provided that they all loaded it in shared mode.
When writing to memory, a process loads a cache line into its cache in \textit{exclusive} mode, and makes the change to its own cached copy (gradually flushing the change downward to all lower levels of the memory hierarchy).
Loading a cache line in exclusive mode also causes any copies of that cache line in other process' caches to be \textit{invalidated}.
Whenever a process $p$ loads a cache line that was invalidated since $p$ last accessed it, $p$ will experience a last-level cache miss.

\paragraph{Non-uniform memory architectures (NUMAs)}

Many modern systems have non-uniform memory architectures, in which different parts of memory can have drastically different access times for different processes.
We take a simplified view of NUMAs that captures the most important costs.
Processes run on one or more \textit{sockets}.
All processes on a socket share the same last-level cache.
Writes by a process do \textbf{not} cause cache invalidations for any processes that share the same cache (but do cause cache invalidations for processes on other sockets).

As an example, consider two processes $p$ and $q$ that are on the same socket (and, hence, share the same last-level cache).
Suppose $q$ loads a cache line, then $p$ writes to it, and then $q$ loads it again.
Since $p$ and $q$ share the same last-level cache, when $p$ performs its write, it simply modifies $q$'s copy that is already in the last-level cache, and does \textbf{not} invalidate it.
However, $p$'s write will still invalidate any copies of the cache line in the higher level caches, and in the last-level caches of other sockets.
Thus, the next time $q$ accesses this cache line, it will use the copy in its last-level cache.
In contrast, if $p$ and $q$ were on different sockets, then $q$'s second load would have to retrieve the cache line from main memory.

\paragraph{Hardware transactional memory}

%

Some of the related work that we discuss relies on hardware transactional memory (HTM), which allows a programmer to execute arbitrary blocks of code atomically as transactions.
Each transaction either \textit{commits} and appears to take effect instantaneously, or \textit{aborts} and has no effect on shared memory.
The set of memory locations read (resp., written) by a transaction is called its \textit{read-set} (resp., \textit{write-set}).
The \textit{data-set} of a transaction is the union of its read-set and write-set.
If the write-set of a transaction intersects the data-set of another transaction, then the two transactions are said to \textit{conflict}.
When two transactions conflict, one of them must abort to ensure consistency.

HTM support has appeared in several commercially available processors produced by Intel and IBM.
This support consists of instructions for starting, committing and aborting transactions, and some platform specific offerings.
In these HTM implementations, transactions can abort not only because of conflicts, but also for other spurious reasons.
These HTM implementations are \textit{best-effort}, which means they offer no guarantee that any transaction will ever commit.
We consider Intel's implementation of HTM.

\end{thesisnot}

\section{Related work} \label{sec-debra-related}

There are many existing techniques for reclaiming memory in lock-free data structures.
We give a detailed survey of the literature, and identify significant problems with some of the most widely used memory reclamation algorithms, and some of the most recent ones.
These problems are poorly understood, and make these reclamation algorithms unsuitable for use with a large class of lock-free algorithms.

\paragraph{Reference Counting (RC).}
RC augments each \record\ $o$ with a counter that records the number of pointers that processes, entry points and \record s have to $r$.
A \record\ can safely be freed once its reference count becomes zero.
Reference counts are updated every time a pointer to a \record\ is created or destroyed.
Naturally, a process must first read a pointer to reach a \record\ before it can increment the \record's reference counter.
This window between when a \record\ is reached and when its reference counter is updated reveals the main challenge in designing a RC scheme: the reference count of a freed \record\ must not be accessed. 
However, the reference count of a retired \record\ can be accessed.

Detlefs et~al.~\cite{Detlefs2002} introduced lock-free reference counting (LFRC), which is applicable to arbitrary lock-free data structures.
LFRC uses the double compare-and-swap (DCAS) synchronization primitive, which atomically modifies two arbitrary words in memory, to change the reference count of a \record\ only if a certain pointer still points to it.
DCAS is not natively available in modern hardware, but it can be implemented from CAS~\cite{Harris:2002}. 
Herlihy et~al.~\cite{Herlihy2005} subsequently improved LFRC to single-word lock-free reference counting (SLFRC), which uses single-word CAS instead of DCAS.
To prevent the reference count of a freed \record\ from being accessed, the implementation of SLFRC uses a variant of Hazard Pointers (described below) to prevent \record s from being freed until any pending accesses have finished.
Lee \cite{Lee2010} developed a distributed reference counting scheme from fetch-and-increment and swap.
Each node contains several limited reference counts, and 
the true reference count for a node is distributed between itself and its parent.
The scheme was developed for in-trees (in which each node only has a pointer to its parent), but it may be generalizable. 

RC requires extra memory for \textit{each \record} and it cannot reclaim \record s whose pointers form a cycle (since their reference counts will never drop to zero).
Manually breaking cycles to allow reclamation requires knowledge of the data structure and adds more overhead.
RC has high overhead, since following a pointer involves incrementing, and later decrementing, its reference count, which is expensive.
Experiments confirm that RC is less efficient than other techniques \cite{Hart2007}.

\paragraph{Hazard Pointers (HPs).}
Michael introduced HPs \cite{Michael2004}, and provided a wait-free implementation from atomic read/write registers.
(Herlihy et~al.~\cite{Herlihy2005} independently developed another version of HPs called Pass-the-Buck (PTB), providing a lock-free implementation from CAS, and a wait-free implementation from double-wide CAS.) 
HPs track which \record s might be accessed by each process.
Before a process can access a field of a \record\ $r$, or use a pointer to $r$ as the expected value for a CAS, it must first acquire a hazard pointer to $r$. 
To correctly use HPs, one must satisfy the following constraint. 
%
Suppose a \record\ $r$ is retired at time $t_r$ and later accessed by a process $p$ at time $t_a$.
Then,
one of $p$'s HPs must continuously point to $r$ from before $t_r$ until after $t_a$.
%
This constraint implies that, after a \record\ $r$ is retired and is not pointed to by any HP, no process can acquire a HP to $r$ until it is freed, allocated again, and inserted back into the data structure.
Therefore, a process can safely free a retired \record\ after scanning all HPs and seeing that none of them point to $r$.

To acquire a HP to $r$, a process first announces the HP by writing a pointer to $r$ in a shared memory location only it can write to. 
This announces to all processes that $r$ might be accessed and cannot safely be freed.
Then, the process verifies that $r$ is in the data structure.
If $r$ is not in the data structure, then the process can behave as if its operation had failed due to contention (typically by 
restarting its operation), without threatening the progress guarantees of the data structure.
%
As we will discuss below, for many data structures, a process cannot easily tell with certainty whether a record is in the data structure.
Such data structures are modified in ad-hoc ways so that operations restart whenever they cannot tell whether a record is in the data structure.
This requires reproving the data structures' progress guarantees (a subtlety that has been missed by many).


On modern Intel and AMD systems, a \textit{memory barrier} must be issued immediately after a HP is announced to ensure that the announcement is immediately flushed to main memory, where it can be seen by other processes.
Otherwise, 
a HP announcement might be delayed so that a process performing reclamation will miss it, and erroneously free the \record\ it protects.
Memory barriers are costly, and this introduces significant overhead.

Many lock-free algorithms require only a small constant number $k$ of HPs per process, since a HP can be released once a process will no longer access the \record\ to which it points during an operation.
Scanning the HPs of all processes takes $\Theta(nk)$ steps, so doing this each time a \record\ is retired would be costly.
However, with a small modification, the expected amortized cost to retire an object is $O(1)$. 
Each process maintains a local collection of \record s that it has removed from the data structure.
(Note that each record is only removed by one process.)
When a collection contains $nk+\Omega(nk)$ objects, the process creates a hash table $T$ containing every HP, and, for each object $o$ in its collection, checks whether $o$ is in $T$.
If not, $o$ is freed. 
Since there are at most $nk$ \record s in $T$,
$\Omega(nk)$ \record s can be freed. 
Thus, the number of \record s waiting to be freed is $O(kn^2)$. 
To obtain good performance, one typically chooses a fairly large constant for the $\Omega(kn^2)$ term.

Aghazadeh et~al.~\cite{Aghazadeh2014} introduced an improved version of HPs with a worst case constant time procedure for scanning HPs each time a \record\ is retired.
Their algorithm maintains two queues of length $nk$ for each process.
These queues are used to incrementally scan HPs as \record s are retired.
The algorithm adds a limited type of reference count to each \record\ that tracks the number of incoming references from one of the queues.
Note that $\Theta(\log(nk))$ bits are reserved in each \record\ for the reference count.

Recall that, with traditional HPs, a process must issue a costly memory barrier immediately after announcing a HP.
Dice et~al.~\cite{Dice2016} recently introduced three techniques for implementing HPs more efficiently by eliminating these frequent barriers. 
The first technique harnesses the \textit{memory protection} mechanism of modern operating systems.
Whenever a process is about to scan all hazard pointers to reclaim memory, it first \textit{write-protects} the memory pages that contain all hazard pointers.
Enabling write-protection on a page acts like a global memory barrier that causes all processes to flush any pending writes to the page before it is write-protected.
Thus, from the perspective of the process performing memory reclamation, it is as if all processes had been issuing memory barriers immediately after their HP announcements.
Unfortunately, this technique is \textit{not} lock-free, since a process that crashes during reclamation will cause all processes to block.
Nevertheless, it could be useful for lock-based algorithms that perform searches without acquiring locks.
The second technique exploits an idiosyncrasy of certain x86 architectures to achieve a non-blocking variant of the first technique.
However, the authors stress that this is not a portable solution. 
The third technique is a hardware-assisted mechanism that relies on an extension to current processor architectures suggested by the authors.

\paragraph{Problems with Hazard Pointers.}

A major problem with HPs is that they cannot be used with many lock-free data structures.
Recall that in order to acquire a HP to a record $r$, one must first announce a HP to $r$, then verify that $r$ is not retired.
If one can determine that $r$ is \textit{definitely not retired}, then a HP to $r$ has successfully been acquired.
On the other hand, if one can determine that $r$ is \textit{definitely retired}, then the operation can simply behave as if it failed due to contention (typically restarting).
However, in many data structures, it is not clear how an operation can determine whether a node is \textit{definitely} retired or not retired.
And, if the operation cannot tell for sure which is the case, then behaving as if it failed due to contention can cause the data structure to lose its progress guarantee, as in the following example.

Many data structures with lock-free operations use \textit{marking} to prevent processes from erroneously performing modifications to records just before, or after, they are removed from the data structure.
Specifically, before a record is removed from the data structure, it is marked, and no process is allowed to change a marked node.
Search operations can often traverse marked nodes, and even leave the data structure to traverse some retired nodes, and still succeed (see, e.g.,~\cite{arbel2014concurrent,Brown:2014,BH11,Drachsler2014,Ellen:2010,Ellen2014,Heller2005,Howley:2012,Natarajan:2014,Ramachandran2015,Shafiei:2013}).

As an example, consider a lock-free singly-linked list in which nodes are marked before they are retired, and operations can traverse retired nodes.
Suppose that, while searching this list, a process $p$ has acquired a HP to node $u$ and needs to acquire a HP to the next node, $u'$.
To acquire the HP to $u'$, $p$ reads and announces the pointer to $u'$, and must then \textit{verify that $u'$ is in the list}.

One might initially think of checking whether $u'$ is marked to determine whether it is in the list.
Since nodes are marked before they are retired, a node that is not marked is definitely not retired.
Unfortunately, there are two problems with this approach.
First, in order to check whether $u'$ is marked, $p$ would already need to have a HP to $u'$.
Second, if $p$ were to see that $u'$ \textit{is} marked, then it would learn nothing about whether $u'$ is actually in the list.
(Since nodes are marked \textit{before} they are retired, a process may have marked $u'$ but not yet retired it.)

We can get around the first problem by having $p$ check whether the \textit{previous} node $u$ is marked, instead of checking whether $u'$ is marked.
(Recall that $p$ already has a HP to $u$.)
If $u$ points to $u'$ and is not marked, then $u$ and $u'$ are definitely both in the list.
However, this does not resolve the second problem: if $u$ is marked, we learn nothing about whether $u$ is in the list (and, hence, we learn nothing about whether $u'$ is in the list).
All we have done is reduced the problem of determining whether $u'$ is in the list to the problem of determining whether $u$ is in the list.

To resolve the second problem, one might think of having $p$ continue to move backwards in the list until it reaches a node that is not marked.
(Since nodes do not change after they are marked, if $p$ were to find a node that is not marked, and points to the chain of marked nodes $p$ traversed, then $p$ would know that all of these nodes are in the list.)
However, unless $p$ holds HPs to \textit{every} node it traverses while moving backwards in the list, it will run into the first problem again.
In the worst case, $p$ will have to hold HPs to every node it visits during its search.
Thus, the number of HPs needed by each process can be \textit{arbitrarily large}.
This can introduce significant space and time overhead, and also enables a crashed process to prevent an arbitrarily large number of nodes from being reclaimed.
Additionally, since the list is singly-linked, $p$ will have to remember all of the previous nodes that it visited, which may require changes to the algorithm.

Another possible way to verify that $u'$ is in the list is to recursively start a new search looking for $u'$.
Of course, this new search can encounter the exact same problem at a different node on the way to $u'$.
Additionally, $p$ would require additional HPs to perform this search (without releasing the HPs it currently holds on $u$).
Clearly, this approach would be extremely complex and inefficient.
Moreover, for some data structures, the amortized cost of operations depends on the number of marked nodes they traverse.
Searching again from an entry point can provably lead to an asymptotic increase in the amortized cost of operations~\cite{Ellen2014}.

Given that lots of data structures use HPs in practice, one might ask how they deal with these problems.
In practice, it is common for data structures that use HPs to simply restart an operation whenever a marked node is encountered.
This is usually done without any concern for whether restarting operations will preserve the data structure's progress guarantees.
We argue that this will not always preserve lock-freedom.
(And we suspect that this will \textit{almost always violate} lock-freedom.)
In our example, suppose $p$ sees that $u$ is marked, and restarts its search without being certain that $u$ is actually retired.
If the process that marked $u$ crashed before actually retiring $u$, then after $p$ restarts the search, it will simply encounter $u$ again, see that it is marked, and restart again.
Thus, $p$ will restart its search forever, never making progress.
Other processes can get into similar infinite loops just as easily, so this can cause all processes to block, violating lock-freedom.
It is straightforward to see that restarting in this way will violate lock-freedom for search procedures in many data structures (including~\cite{arbel2014concurrent,Brown:2014,BH11,Drachsler2014,Ellen:2010,Ellen2014,Heller2005,Howley:2012,Natarajan:2014,Ramachandran2015,Shafiei:2013}).

Additionally, HPs introduce complications in data structures with \textit{helping}. 
In many lock-free data structures, whenever a process $p$ is prevented from making progress by another operation $O$ (perhaps because $O$ has marked a node that $p$ would like to modify), $p$ \textit{helps} $O$ by performing the steps necessary to complete $O$ (removing any marked nodes from the data structure), on behalf of the process that started $O$.
This involves accessing several nodes, and modifying some of them.
Of course, before $p$ can access these nodes, it must acquire HPs to them.
As we explained above, in order for $p$ to acquire HPs to these nodes, it must be able to determine that they are definitely not retired.
Consequently, one cannot use helping to resolve the problems described above.

Note that, more generally, these problems occur whenever a process can traverse a pointer from a retired node to another retired node (regardless of whether nodes are marked).
It appears this scenario can occur in any data structure where retired records can point to records that are still in the data structure (which can later be retired), and searches do not help other operations (which means they cannot restart without violating lock-free progress).

Considering the exposition above, it seems likely that precisely determining whether nodes are retired after announcing HPs is approximately as difficult as maintaining a reference count in each node that counts the number of incoming pointers from other nodes in the data structure.

\paragraph{Beware \& Cleanup (B\&C).}
B\&C
was introduced by Gidenstam et~al.~\cite{Gidenstam2009} to allow processes to acquire hazard pointers to \record s that have been retired but not reclaimed.
A limited form of RC is added to HPs to ensure that a retired \record\ is freed only when no other 
\record\ points to it.
The reference count of a \record\ counts incoming pointers from other \record s, but does not count incoming pointers from processes' local memories.
Before reclaiming a \record, a process must verify that its reference count is zero, and that no HP points to it.
Consequently, 
after announcing a HP to a \record\ $r$, to determine whether $r$ is retired, it suffices to check whether its reference count is nonzero. 

Unfortunately, if a data structure allows retired records to point to other retired records, then a retired record's reference count may never be zero.
To address this issue, the authors make the following assumption: ``each [pointer] in a [retired record] that references another [retired record] can be replaced with a reference to [a record that is not retired], with retained semantics for all of the involved threads.''
To use B\&C, one must implement a procedure that takes a retired record and changes it so that it no longer points to any retired record.
Designing such a procedure and proving that it ``retains semantics for all of the involved threads'' is non-trivial.
Additionally, B\&C's algorithm for retiring \record s is extremely complicated, and the technique has significantly higher overhead than HPs.

The authors did not mention the problems with HPs described above.
They stated that operations using HPs would need to restart whenever they encountered a retired \record, but did not consider how such operations would actually determine whether a \record\ is retired.
The goal of the work was simply to \textit{improve performance} for operations that frequently restart.
Nevertheless, their technique does solve the problems with HPs described above.
Regrettably, it does not appear to be practical.


\paragraph{ThreadScan (TS).}
TS is a variant of hazard pointers that avoids making an expensive announcement for each record accessed by treating the private memory of each process as if every pointer it contains is an announcement~\cite{Alistarh:2015}. 
TS was developed independently, at the same time as DEBRA+.
Like DEBRA+, TS uses uses operating system signals to enable threads obtain a progress guarantee.

Each process maintains a single-writer multi-reader \textit{delete buffer} in shared memory that contains the \record s it has removed from the data structure, but has not yet freed.
When process $p$'s buffer becomes sufficiently large, $p$ starts \textit{reclamation} by acquiring a global lock (which prevents other processes from starting reclamation).
It then collects the records in the buffers of all processes, and sends a signal to every other process.
Whenever a process $q$ receives a signal, it scans through its own private memory (stack and registers) 
and \textit{marks} each \record\ that it has a pointer to.
Then, $q$ sends an acknowledgment back to $p$, indicating that it has finished marking \record s.
Process $p$ waits until it receives acknowledgments from all processes, and then frees any \record s in its buffer that are not marked.

The authors claim that TS offers strong progress guarantees under the assumption that: ``the operating system does not starve threads.''
However, this assumption is extremely strong.
It implies that processes cannot fail, which means that TS cannot be used by a lock-free algorithm.
There are two reasons why TS needs this assumption.
First, TS uses a global lock to ensure that only one process is performing reclamation at a time.
Second, a process performing reclamation must wait to receive acknowledgments from all other processes.

Although TS is not fault-tolerant, it is a very attractive option in practice, since it is easy to use, and it appears to be quite efficient.
(Preliminary experiments show that it performs similarly to DEBRA+.)
In order to use TS, one simply invokes a procedure whenever a process has just removed a \record\ from the data structure.
However, TS can only be applied to algorithms that satisfy a set of assumptions.
One of these assumptions is particularly problematic: ``nodes in [a process'] delete buffer have already been removed [from the data structure], and cannot be accessed through shared references, or through references present on the heap.''
Restated using our terminology, this assumption says that records in a process' delete buffer are retired, and cannot be accessed by following pointers from other records (even other retired records).
The authors claim that this assumption ``follows from the definition of concurrent memory reclamation~\cite{Michael2004,Herlihy2005}.'' 
However, it is not clear that this assumption follows from the definitions in those papers. 

Consider the relevant part of the definition in~\cite{Michael2004}:
A hazard pointer to a \record\ must be acquired before the \record\ is retired.
This admits the possibility that a process can hold hazard pointers to (and, hence, safely access) potentially many retired \record s, as long as the necessary hazard pointers were acquired before these \record s were retired.
Thus, it is theoretically possible for a process to follow pointers from retired \record s to other retired \record s (although there are some problems in practice with traversing pointers from retired \record s, as we discussed above). 
Similarly, in the pass the buck algorithm of Herlihy et~al.~\cite{Herlihy2005}, it is safe to access any \record\ that is \textit{injail}, which means it has been allocated since it was last freed.
Since \record s are retired before being freed, any \record\ that is retired but has not yet been freed is still \textit{injail}.
Thus, it is perfectly fine to follow pointers in shared memory from retired \record s to other retired \record s.
Therefore, it would appear that TS's assumption above is strictly stronger than each of these.

\paragraph{Applicability of TS.}
TS's assumptions prevent it from being used with algorithms where a process can traverse a pointer in shared memory from a retired \record\ to another retired \record .
This includes all of the algorithms listed where we discussed the problem with HPs.
We briefly consider what happens if such an algorithm is used with TS.

Suppose a process $p$ has a pointer in its private memory to a node $u$ that another process $q$ has removed from the data structure, and is about to read a pointer in $u$ that points to another node $u'$ which $q$ has also been removed from the data structure.
Then, suppose that, before $p$ reads the pointer in $u$ that it would follow to reach $u'$, $q$ begins reclamation, and signals all processes, including $p$.
This causes all processes to stop what they are doing and help $q$ perform reclamation, by scanning their private memories for any pointers to nodes in $q$'s buffer (including $u$ and $u'$).
If no process has a pointer to $u'$ in its memory, then no process will mark node $u'$, and process $q$ will free $u'$.
This can occur because $p$ has a pointer to $u$ in its private memory, but does have a direct pointer to $u'$.
However, since $u$ points to $u'$, after $p$ finishes helping $q$ perform its reclamation, and resumes its regular algorithm, it will follow the pointer from $u$ to $u'$, performing an illegal access to a freed node.

\paragraph{Dynamic Collect}

Dragojevi{\'c} et~al. \cite{Dragojevic2011} explored how hardware transactional memory (HTM) can be used to easily produce several implementations of a \textit{dynamic collect} object.
A dynamic collect object has four operations: \textit{Register}, \textit{DeRegister}, \textit{Update} and \textit{Collect}.
Intuitively, one can think of a dynamic collect object as a collection of hazard pointers that can increase and decrease in size. 
Register adds a new HP to the dynamic collect object, and DeRegister removes one.
Update sets a HP. 
An invocation of Collect returns the set of HPs that were set before the Collect began (and were not removed during the Collect), and possibly some others. 
Similarly to HPs, it is not clear how one could use dynamic collect to reclaim memory for a lock-free data structure wherein processes can traverse pointers from retired records to other retired records.

One interesting observation made by the paper is that, in RC schemes, if a process traverses several records inside a transaction, then some increments and decrements of reference counts can be elided.
We explain with an example.
Let $r_1$, $r_2$ and $r_3$ be records in a data structure.
Suppose a process executing in a transaction follows a pointer from $r_1$ to $r_2$, and a pointer from $r_2$ to $r_3$.
Observe that it is not necessary to increment or decrement $r_2$'s reference count, since it would be incremented when the process follows the pointer from $r_1$ to $r_2$, and decremented when the process follows the pointer from $r_2$ to $r_3$ (and the atomicity of transactions guarantees that neither change will be visible on its own).
In general, if a process follows a chain of pointers inside a transaction, only the reference counts of the first and last records must be updated. 
The authors described how to efficiently traverse large linked lists by splitting a traversal into many small transactions, and using this technique to eliminate most of the overhead of reference counting.

\paragraph{StackTrack (ST).}
Alistarh et~al.~\cite{Alistarh2014} introduced an algorithm called StackTrack, which relies on the cache coherence protocol of the processor to detect conflicts between a process that frees a \record\ and a process that accesses the \record.
The key idea is to execute each operation of the lock-free data structure in a transaction, and use the implementation of HTM to automatically monitor all pointers stored in the private memory of processes without having to explicitly announce pointers to them before they are accessed.
If a transaction accesses a \record\ which is freed during the transaction, then the HTM system will simply abort the transaction, instead of causing the failure of the entire system.

To decrease the probability of aborting transactions, each operation is split into many small transactions called \textit{segments}.
This takes advantage of the fact that lock-free algorithms do \textit{not} depend on transactions for atomicity.
Segments are executed in order, with each segment being repeatedly attempted until it either commits or has aborted $k$ times.
The size of segments is adjusted dynamically based on how frequently segments commit or abort.
If segments frequently abort, then they are made smaller (which increases overhead, but makes them more likely to commit), and vice versa.
If a segment aborts $k$ times, then a non-transactional fallback code path for the segment is executed instead.
This fallback path uses HPs.
%
A process $p$ executing a segment on the fallback path may need to access some records that it obtained pointers to while executing its previous segment.
In order to access these records, $p$ must know that they have not been freed.
Thus, $p$ must already have HPs to these records when it begins executing on the fallback path.
Consequently, at the end of each segment executed \textit{in a transaction} by a process $p$, all pointers in $p$'s private memory are announced as HPs.

Any time a process wants to free a \record\ $r$, it must first verify that no HP points to $r$.
Each time a process removes a record $r$ from the data structure, it places $r$ in a local list.
If the size of the list exceeds a predefined threshold, then the process invokes a procedure called \textit{ScanAndFree}, which iterates over each record $r$ in the list, and frees $r$ if it is not announced.
%
ST requires a programmer to insert code before and after each operation, whenever a record is retired, and after every few lines of lock-free data structure code.

\paragraph{A Problem with StackTrack.}

Although no such constraint is stated in the paper, ST cannot be applied to any data structure in which an operation traverses a pointer from a retired \record\ to another retired \record 
~\cite{Matveev2014}.
This includes all of the data structures mentioned above where we discussed the problems with HPs.

We briefly explain what happens when ST is used to reclaim memory for such a data structure.
Consider a simple list consisting three nodes, $A$, $B$ and $C$, where $A$ points to $B$ and $B$ points to $C$.
For simplicity, we assume the list supports an operation that atomically deletes two consecutive nodes. 
Suppose process $p$ is searching the list for $C.key$, and process $q$ is concurrently removing $B$ and $C$.
Process $p$ starts a transaction, obtains a pointer to $B$ by reading $A.next$, announces $B$, and commits the transaction.
Then, process $q$ starts a transaction, changes $A.next$ to NULL (to remove $B$ and $C$ from the list), commits the transaction, and adds $B$ and $C$ to its list of removed records.
Next, $q$ invokes \textit{ScanAndFree} and sees that $p$ has announced $B$, so $q$ does not free $B$.
However, $p$ has not announced $C$, so $q$ frees $C$.
Then, $p$ starts a new transaction and reads $B.next$, obtaining a pointer to $C$, and then follows that pointer, which causes the transaction to abort.
This transaction will abort every time it is retried.
Consequently, the data structure operation cannot make progress on the fast path.
Furthermore, as we described above, the HP fallback path cannot accommodate data structures where operations traverse pointers from a retired \record\ to another retired \record. 

\paragraph{Epochs.}
A process is in a \textit{quiescent state} whenever it does not have a pointer to any \record\ in the data structure.
A grace period is any time interval during which every process has a point when it is in a quiescent state.
%
Fraser \cite{Fraser2004} described epoch based reclamation (EBR), which 
assumes that a process is in a quiescent state between its successive data structure operations.
More specifically, EBR can be applied only if processes cannot save pointers read during an operation and access them during a later operation.
\begin{fullver}
We expand on the brief description of EBR given at the beginning of the chapter.
\end{fullver}
\begin{shortver}
We expand on the brief description of EBR given above.
\end{shortver} 
EBR uses a single global counter, which records the current \textit{epoch}, and an announce array.
Each data structure operation first reads and announces the current epoch $\epsilon$, and then checks whether all processes have announced the current epoch.
If so, it increments the current epoch using CAS.
The key observation is that the period of time starting from when the epoch was changed from $\epsilon-2$ to $\epsilon-1$ until it was changed from $\epsilon-1$ to $\epsilon$ is a grace period (since each process announced a new value, and, hence, started a new operation).
So, any \record s retired in epoch $\epsilon-2$ can safely be freed in epoch $\epsilon$.
Whenever a \record\ is retired in epoch $\epsilon$, it is appended to a limbo bag for that epoch.
It is sufficient to maintain three limbo bags (for epochs $\epsilon$, $\epsilon-1$ and $\epsilon-2$, respectively).
Whenever the epoch changes, every \record\ in the oldest limbo bag is freed, and that limbo bag becomes the limbo bag for the current epoch.

Since EBR only introduces a small amount of overhead at the beginning of each \textit{operation}, it is significantly more efficient than HPs, which requires costly synchronization \textit{each time a new \record\ is accessed}.
The penalty for writing to memory only at the beginning of each operation, rather than each time a new \record\ is accessed, is that processes have little information about which \record s might be accessed by a process that is suspected to have crashed.
Consequently, EBR is not fault tolerant.


Quiescent state-based reclamation (QSBR)~\cite{McKenney:1998} is a generalization of EBR that can be used with data structures where processes can save pointers read during an operation and access them during a later operation.
However, to use QSBR, one must manually identify times when individual processes are quiescent.

\paragraph{Drop the Anchor (DTA).}
Braginsky, Kogan and Petrank \cite{Braginsky2013} introduced DTA, a specialized technique for singly-linked lists, which explores a middle ground between HPs and EBR.
Instead of acquiring a HP each time a pointer to a node is read, a HP is acquired only once for every $c$ pointers read.
When a HP to a node $u$ is acquired, it prevents other processes from reclaiming $u$ and the next $c-1$ nodes currently in the list.
(It also prevents other processes from reclaiming any nodes that are inserted amongst these nodes.)
Suppose a process $q$ performs $s$ operations without seeing any progress by another process $p$.
Then, $q$ will cut all nodes that $p$ might access out of the list, replacing them with new copies
It will also mark the old nodes so that $p$ can tell what has happened.
If $p$ has crashed, then the nodes that $q$ cuts out of the list can never be freed.
However, if $p$ has not crashed, then it will eventually see what has happened, and attempt to free these marked nodes.
Observe that memory reclamation can continue in the list regardless of whether $p$ has crashed.
Consequently, DTA is fault tolerant.

DTA has been shown to be efficient \cite{Alistarh2014,Braginsky2013}.
However, it is not clear how it could be extended to work for other data structures.
Additionally, DTA needs to be integrated with the mechanism for synchronizing updates to the linked list, because a sequence of nodes can be cut out of the list concurrently with other updates.

In the worst case, the number of retired nodes that cannot be freed is $\Omega(scn^2)$.
To see why, consider the following.
Let $p$ be a process with a HP at a node $u_1$ that allows it to access nodes $u_1, ..., u_c$.
Suppose each process other inserts $s$ nodes between $u_1$ and $u_c$ before $p$ performs another step. 
Observe that $p$'s next step might access any of the $\Omega(scn)$ nodes starting at $u_1$ and ending at $u_c$.
Thus, a process $q$ that suspects $p$ has crashed will cut all of these nodes out of the list.
None of these nodes can be freed if $p$ crashes.
If this is repeated for each process $p \neq q$, then there will be $\Omega(scn^2)$ nodes that cannot be reclaimed.

\medskip
\noindent\textbf{QSense (QS).}
%
Recently, Balmau et~al.~\cite{Balmau2016} introduced QS, another algorithm that combines HPs and EBR.
Like DTA, QS adds fault-tolerance to EBR by using HPs.
Like the accelerated implementations of HPs in~\cite{Dice2016}, the performance benefit of QS over HPs comes from a reduction in the overhead of memory barriers issued to ensure that HP announcements are visible to all threads.
At a high level, QS uses two execution paths: an EBR-based fast path and a HP-based slow path.
As long as processes continue to make progress, they continue to use the fast path.
If a process has not made progress for a sufficiently long time, then all processes switch to the slow path.
To guarantee that nodes are not erroneously freed when the algorithm switches from the fast path to the slow path, the fast path and slow path both acquire HPs.
On the fast path, EBR is effectively used to reduce the cost of reclamation by eliminating the need to scan HPs to determine whether nodes can be freed.

In order to reduce the overhead of issuing memory barriers for the HPs acquired in QS, the authors make the following observation: In a modern operating system, running on an x86/64 architecture, whenever a process experiences a context switch, the kernel code for performing a context switch 
issues at least one memory barrier.
Suppose a process $p$ announces a HP at time $t$, and does \textit{not} perform a memory barrier after announcing the HP.
Additionally, suppose another process $q$ begins scanning HPs (to perform reclamation) at time $t' > t$.
If $p$ experiences a context switch between $t$ and $t'$, then $q$ will see $p$'s HP announcement, as if $p$ had issued a memory barrier.
(This is somewhat similar to the HP schemes of Dice et al.~\cite{Dice2016} discussed above, which also harness operating system and/or hardware primitives to eliminate memory barriers.)

The authors introduce \textit{rooster processes} to trigger context switches for all processes at regular intervals.
Each processor has a rooster process pinned to it that sleeps for some fixed length of time $T$, then wakes up (forcing a context switch), then immediately sleeps again.
Whenever a process wants to reclaim a node $u$ that was retired at time $t$, it waits until time $t+T+\epsilon$ (for some small $\epsilon > 0$), by which point a rooster process should have woken up, forcing a context switch and guaranteeing that the reclaiming process can see any HPs announced before the node was removed from the data structure.
The $\epsilon$ term above is necessary because, in real systems, when a process requests to sleep for $T$ time units, it may sleep longer.
If a rooster process sleeps for more than $t+T+\epsilon$ time, then its failure to trigger a timely context switch might cause a reclaiming process to miss a HP and erroneously free a node that is still in use.
Thus, \textit{QS only works under the following assumption}: a bound on $\epsilon$ must be known and rooster processes never fail.
Consequently, QS does not work in a fully asynchronous system.
(In comparison, the TS algorithm also makes timing assumptions, but it only loses its \textit{progress} guarantee in a fully asynchronous system.)

QS switches from the slow path to the fast path only when every process has completed an operation since it last switched to the slow path.
Consequently, if a process crashes either before or while processes are executing on the slow path, then all processes remain on the slow path forever (where they cannot use EBR to reclaim memory).
This is not true for DEBRA+, which allows EBR to continue, even if a process has crashed.
Furthermore, since QS uses HPs, it cannot be used with data structures where operations traverse pointers from a retired record to another retired record (as we described above).

\medskip
\noindent\textbf{Optimistic Access (OA).}
Cohen and Petrank~\cite{Cohen2015} introduced an approach where algorithms optimistically access parts of memory that might have already been reclaimed, and check after each read whether this was the case.
(Their approach was developed independently, at the same time as DEBRA+.)
Crucially, OA relies on processes being able to access reclaimed memory without causing the system to crash.
Consequently, an algorithm that uses OA must either (a) never release memory to the operating system, or (b) trap segmentation fault and bus fault signals and ignore them.
%
If option (a) is used, then OA has the same downsides as OPs.
On the other hand, if option (b) is used, then one must cope with the downsides of trapping segmentation and bus faults.
In particular, if an algorithm contains bugs that cause segmentation or bus faults, then option (b) makes it significantly more difficult to identify these bugs, because one cannot easily distinguish between faults caused by a program bug and faults caused by accesses to reclaimed memory.
Such algorithms cannot be used in software that already traps segmentation or bus faults, including many common debugging tools.
(Incidentally, 
like DEBRA+, option (b) is lock-free only if the operating system's signaling mechanism is lock-free.)

OA can be used only with algorithms that appear in a \textit{normalized} form.
At a high level, an operation in a normalized algorithm proceeds in three phases: CAS generation, CAS execution and wrap-up.
In the CAS generation phase, the operation reads shared memory and constructs a sequence of CAS steps.
These CAS steps are performed in the CAS execution phase.
Any additional processing is performed in the wrap-up phase.
In the CAS execution and wrap-up phases, the operation can be helped by other processes.
As we will see, the class of algorithms in lock-free normalized form is quite similar to the class of algorithms that can use DEBRA+ in a straightforward way.

At a high-level, OA works as follows.
Each process $p$ has a \textit{warning} bit that is cleared whenever $p$ starts a new operation, and is set whenever any process begins reclaiming memory.
After $p$ performs a read from shared memory, it checks whether its warning bit is set, and, if so, jumps back to a safe checkpoint earlier in the operation (intuitively restarting the operation).
Unlike reads, CAS operations cannot be optimistically performed on reclaimed memory, since this could cause data corruption.
Thus, \textit{before} performing CAS on any \record, a process first announces a HP to ensure that another process does not reclaim the \record\ (and possibly does the same for the old value and new value for the CAS), and then checks whether its warning bit is set.
If so, it releases its HP and jumps back to a safe checkpoint.
Whenever $p$ retires a \record, it places it in a local buffer.
If this local buffer becomes sufficiently large, then $p$ performs reclamation.
To perform reclamation, $p$ simply sets the warning bits of all processes, and then frees every \record\ in its local buffer that is not pointed to by a HP.

Before one can use OA to reclaim memory for a lock-free algorithm, one must first transform the algorithm into normalized lock-free form, then replace each shared memory read or CAS with a small procedure.
Each shared memory read is replaced with: two reads, a branch, a possible write, and a possible jump.
Each shared memory write or CAS is replaced with: four to seven writes, 
one read, one branch, a CAS, a memory fence, three possible bit-masking operations, and a possible jump.

Conceptually, OA is somewhat simpler than DEBRA+.
However, in practice, OA would likely be less efficient, and more time consuming to apply to complex data structures, since it requires code modifications for each read, write and CAS on shared memory, and for each \record\ retired.
It is also quite likely to be less efficient, for this reason.
In a subsequent paper, Cohen and Petrank presented an extended version of OA called automatic optimistic access (AOA).
AOA uses garbage collection techniques to eliminate the need for a procedure to be invoked whenever a record is retired~\cite{Cohen2015AOA}.

\begin{figure*}[tb]
    \setlength\tabcolsep{4pt}
	\vspace{-2mm}
    \small
	\centering
	\begin{tabular}{|l|c|c|c|c|c|c|c|c|c|c|c|c|}
	\hline
	Necessary code modifications 			   & RC & HP & B\&C & TS & ST & EBR & DTA & QS & OA & DEBRA & DEBRA+ \\
	\hline
	\hspace{2mm} per accessed \record\ 	& \cmark & \cmark & \cmark & & \cmark & & \cmark & \cmark & \cmark & & \\
	\hspace{2mm} per operation 	 & & & & \cmark & & \cmark & \cmark & \cmark & & \cmark & \cmark \\
	\hspace{2mm} per retired \record\ 	& & \cmark & \cmark & \cmark & \cmark & \cmark & \cmark & \cmark & \cmark & \cmark & \cmark \\
	\hspace{2mm} other 			& a & b & a & & b, c & & d & b & e & & f \\
	\hline
    Special timing assumptions & & & & For progress & & & & For correctness & & & \\
    \hline
    Fault tolerant & \cmark & \cmark & \cmark & & \cmark & & \cmark & \cmark & \cmark & & \cmark \\
    \hline
    \begin{tabular}{@{}l@{}}Termination of memory\\reclamation procedures\end{tabular} & L & W & L & Blocking & L & L/W & L & L$_{rooster}$ & L & W & W$_{sig}$ \\
    \hline
	\begin{tabular}{@{}l@{}}Can traverse pointer from\\retired \record\ to retired \record\end{tabular} & \cmark & & \cmark & & & \cmark & \cmark & & \cmark & \cmark & \cmark \\
	\hline
	\end{tabular}
	\vspace{-2mm}
	\caption{
	Summary of reclamation schemes.
	\textbf{Other code modifications:} (a) break cycles in pointers;
	(b) write recovery code for when a process fails to acquire a HP;
	(c) insert transaction \textit{checkpoints} after every few lines of code;
	(d) integrate crash recovery with the lock-free data structure's synchronization mechanisms (and only works for lists);
	(e) transform lock-free algorithm into normalized form, and then instrument every read, write and CAS on shared memory;
	(f) write crash recovery code (which is trivial for many data structures).
    \textbf{Termination of memory reclamation procedures:} (L) lock-free; (L$_{rooster}$) lock-free if rooster processes cannot crash; (W) wait-free; (W$_{sig}$) wait-free if the operating system's signaling mechanism is wait-free
	}
	\label{fig-related}
	\vspace{-2mm}
\end{figure*}

\medskip
\noindent\textbf{Applying each technique.}
The effort required to apply these techniques varies widely.
See Figure~\ref{fig-related} for a summary.


\section{DEBRA: Distributed Epoch Based Reclamation} 
\label{sec-technique}

\begin{figure*}
\lstset{escapechar=@,style=customc}
\centering
\begin{minipage}{0.44\textwidth}
\centering
(Applying DEBRA)
\begin{lstlisting}[frame=single]
Value search(Key key) {
+   leaveQstate();

    Node *node = root;
    
    while (!node.isLeaf()) {

        if (key < node->key)) {
            node = node->left;
        } else {
            node = node->right;
        }






    }
    if (key == node->key) {
+       Value result = node->value;
+       enterQstate();
        return result;
    }
+   enterQstate();
    return NO_VALUE;
}
\end{lstlisting}
\end{minipage}
\hspace{3mm}
\begin{minipage}{0.48\textwidth}
\centering
(Applying hazard pointers)
\begin{lstlisting}[frame=single]
Value search(Key key) {
+   announce root
+   if (root is retired) restart search
    Node *node = root;
+   Node *prev = NULL;
    while (!node.isLeaf()) {
+       prev = node;
        if (key < node->key)) {
            node = node->left;
        } else {
            node = node->right;
        }
+       announce node
+       if (node is retired) {
+           release hazard pointer to prev
+           restart search
+       }
+       release hazard pointer to prev
    }
    if (key == node->key) {
+       Value result = node->value;
+       release hazard pointer to node
        return result;
    }
+   release hazard pointer to node
    return NO_VALUE;
}
\end{lstlisting}
\end{minipage}
\caption{Applying DEBRA and HPs to a search in a binary search tree. (+) denotes a new line. 
}
\label{fig-using-debra-vs-hp}
\end{figure*}

In this section, we present DEBRA, a distributed version of EBR with numerous advantages over classical EBR.

First, DEBRA supports a sort of \textit{partial fault tolerance}.
Consider an execution of a data structure that uses EBR.
Observe that a process that sleeps for a long time will delay reclamation for all other processes, \textit{even if it is not currently executing an operation on the data structure}.
In DEBRA, a process can prevent other operations from reclaiming memory \textbf{only} if it is currently executing an operation on the data structure.
Thus, if a process sleeps for a long time or crashes while it is not executing an operation on the data structure, other processes will continue to reclaim memory as usual.
This can have a significant impact in real applications, where operations on a data structure may represent a small part of the execution.
It also makes it possible to terminate some of the processes operating on a data structure, or reassign them to different tasks, without permanently halting reclamation for all other processes.

Second, recall that in EBR, each time a process begins a new operation, it reads the epoch announcements of all processes.
This can be expensive, especially on NUMA systems, where reading the epoch announcements of processes on other sockets is likely to incur extremely expensive last-level cache misses.
In DEBRA, we read the epoch announcements of all processes \textit{incrementally} over many operations.
This can slightly delay the reclamation of some records, but it dramatically reduces the overall cost of reading epoch announcements.

Third, instead of having all processes synchronize on shared epoch bags, each process has its own local epoch bags, and reclamation proceeds independently for each process.
Additionally, these bags are carefully optimized for good cache performance and extremely low overhead.

\paragraph{Using DEBRA}

DEBRA provides four operations:
\leaveq$()$, \enterq$()$, \retire$(r)$ and \isq$()$, where $r$ is a record.
Each of these operations takes $O(1)$ steps in the worst-case.
Let $T$ be a lock-free data structure.
To use DEBRA, $T$ simply invokes \leaveq\ at the beginning of each operation, \enterq\ at the end of each operation, and \retire$(r)$ each time a record $r$ is retired (i.e., removed from $T$).
Like EBR, DEBRA assumes that a process does not hold a pointer to any record between successive operations on $T$.
Each process alternates invocations of \leaveq\ and \enterq, beginning with an invocation of \leaveq. 
Each process is said to be \textit{quiescent} initially and after invoking \enterq, and is said to be \textit{non-quiescent} after invoking \leaveq.
An invocation of \isq\ by a process returns true if it is quiescent, and false otherwise. 
Figure~\ref{fig-using-debra-vs-hp} is an example of DEBRA applied to code for searching a lock-free binary search tree.
For comparison, it also shows how HPs could be applied \textit{\textbf{if}} one could determine whether a node is retired.
Note that determining whether node is \textit{retired} (and not just \textit{marked}) can be extremely difficult, as discussed in Section~\ref{sec-debra-related} (where we discussed problems with HPs).

\paragraph{Implementation}

\begin{figure}[th!]
\lstset{escapechar=@,style=customc}
\begin{lstlisting}[frame=single]
process local variables:
    long pid;                           // process id
    long checkNext;                     // the next process whose announcement should be checked
    blockbag *bags[0..2];               // limbo bags for the last three epochs
    blockbag *currentBag;               // pointer to the limbo bag for the current epoch
    long index;                         // index of currentBag in bags[0..2]
    long opsSinceCheck;                 // # ops performed since checking another process' announcement
shared variables:
    long epoch;                         // current epoch
    long announce[n];                   // per-process announced epoch and quiescent bit
    objectpool *pool;                   // pointer to object pool

bool getQuiescentBit(long otherPid)      { return announce[otherPid] & 1; }
void setQuiescentBitTrue(long otherPid)  { announce[otherPid] = announce[otherPid] | 1; }
void setQuiescentBitFalse(long otherPid) { announce[otherPid] = announce[otherPid] & ~1; }
bool isEqual(long readEpoch, long announcement) {
    return readEpoch == (announcement & ~1); // compare read epoch to epoch-bits from announcement
}

void retire(record *p) { currentBag->add(p); }
bool isQuiescent()     { return getQuiescentBit(pid); }
void enterQstate()     { setQuiescentBitTrue(pid); }
bool leaveQstate() {
    bool result = false;
    long readEpoch = epoch;
    if (!isEqual(readEpoch, announce[pid])) { // our announcement differs from the current epoch
        opsSinceCheck = checkNext = 0;    // we are now scanning announcements for a new epoch
        rotateAndReclaim();
        result = true;                    // result indicates that we changed our announcement
    }
    // incrementally scan all announcements
    if (++opsSinceCheck >= CHECK_THRESH) {
        opsSinceCheck = 0;
        long other = checkNext % n;
        if (isEqual(readEpoch, announce[other]) || getQuiescentBit(other)) {
            long c = ++checkNext;
            if (c >= n && c >= INCR_THRESH) { // if we scanned every announcement
                CAS(&epoch, readEpoch, readEpoch+2);
    }   }   }
    announce[pid] = readEpoch;          // announce new epoch with quiescent bit = false
    return result;
}
void rotateAndReclaim() { // rotate limbo bags and reclaim records retired two epochs ago
    index = (index+1) % 3;              // compute index of oldest limbo bag
    currentBag = bags[index];           // reuse the oldest libmo bag as the new currentBag
    pool->moveFullBlocks(currentBag);   // move all full blocks to the pool
}
\end{lstlisting}
\caption{C++ style pseudocode for DEBRA, where $n$ is the number of processes, INCR\_THRESH is the minimum number of times a process must invoke \leaveq\ before it can increment the epoch, and CHECK\_THRESH is the number of times a process must invoke \leaveq\ before it will check the epoch announcement of another process.}
\label{fig-debra}
\end{figure}

C++ style psuedocode for DEBRA appears in Figure~\ref{fig-debra}.
Each process $p$ has three limbo bags, denoted $bag_0$, $bag_1$ and $bag_2$, which contain records that it removed from the data structure.
At any point, one of these bags is designated as $p$'s limbo bag for the current epoch, and is pointed to by a local variable \textit{currentBag}.
Whenever $p$ removes a record from the data structure, it simply adds it to \textit{currentBag}.
Each process has a \textit{quiescent bit}, which indicates whether the process is currently quiescent.
The only thing $p$ does when it enters a quiescent state is set its quiescent bit.
Whenever $p$ \textit{leaves} a quiescent state, it reads the current epoch $e$ and announces it in \textit{announce}$_p$.
If this changes the value of \textit{announce}$_p$, then the contents of the oldest limbo bag can be reused or freed.
In this case, $p$ changes \textit{currentBag} to point to the oldest limbo bag, and then moves the contents of \textit{currentBag} to an object pool. 
Next, $p$ attempts to determine whether the epoch can be advanced, which is the case if each process is either quiescent or has announced $e$.
To do this efficiently, $p$ checks the announcements and quiescent bits of all processes \textit{incrementally}, reading one announcement and one quiescent bit in each \leaveq\ operation.
Process $p$ repeatedly checks the announcement and quiescent bit of the same process $q$ in each of its \leaveq\ operations, until $q$ either announces the current epoch or becomes quiescent, or until the epoch changes.
A local variable \textit{checkNext} keeps track of the next process whose announcement should be checked.
Once \textit{checkNext} is $n$, $p$ performs a CAS to increment the current epoch.


\paragraph{Correctness}

DEBRA reclaims a record only when no process has a pointer to it:
Suppose $p$ places a record $r$ in limbo bag $b$ at time $t_1$, and moves $r$ from $b$ to the pool at time $t_2$.
Assume, to obtain a contradiction, that a process $q$ has a pointer to $r$ at time $t_2$.
At time $t_1$, $b$ is $p$'s current limbo bag, and just before time $t_2$, $b$ is changed from being $p$'s oldest limbo bag to being $p$'s current limbo bag, again.
Thus, \textit{currentBag} must be changed at least three times between $t_1$ and $t_2$.
Since $p$ changes \textit{currentBag} only in an invocation of \leaveq\ that changes \textit{announce}$_p$, $p$ must perform at least three such invocations between $t_1$ and $t_2$.
The current epoch must change between any pair of invocations of \leaveq\ that change \textit{announce}$_p$, so the current epoch must change at least twice between $t_1$ and $t_2$.
Consider two consecutive changes of the current epoch, from $e$ to $e'$ and from $e'$ to $e''$.
At some point between these two changes, $q$ must either be quiescent or have announced $e'$.
Process $q$ must announce $e'$ after reading the current epoch and seeing $e'$, before the current epoch changes from $e'$ to $e''$.
Thus, $q$ must announce $e'$ after the current epoch changes from $e$ to $e'$, and before it changes from $e'$ to $e''$.
Since $q$ can only announce an epoch when it is in a quiescent state, $q$ must therefore be quiescent at some point between $t_1$ and $t_2$.
This means $q$ must obtain its pointer to $r$ by following pointers from an entry point after it was quiescent, which is after $t_1$.
However, $r$ is removed from the data structure before $t_1$, and, hence, it is no longer reachable by following pointers from an entry point.
This is a contradiction.

\paragraph{Object pool}

The object pool shared by all processes is implemented as a collection of $n$ \textit{pool bags} (one per process) and one shared bag.
Whenever a process moves a record to the pool, it places the record in its pool bag.
If its pool bag is too large, it moves some records to the shared bag.
Whenever a process wants to allocate a record, it first attempts to remove one from its pool bag.
If its pool bag is empty, it attempts to take some records from the shared bag.
If the shared bag is empty, the process will allocate some new records and place them in its pool bag.

\paragraph{Block bags}

For efficiency, each pool bag and limbo bag is implemented as a \textit{blockbag}, which is a singly-linked list of \textit{blocks}.
Each block contains a \textit{next} pointer and up to $B$ records.
(In our experiments, $B = 256$.)
The head block in a blockbag always contains fewer than $B$ records, and every subsequent block contains exactly $B$ records.
With this invariant, it is straightforward to design constant time operations to add and remove records in a blockbag, and to move all full blocks from one blockbag to another.
This allows a process to move all full blocks of records in its oldest limbo bag to the pool highly efficiently after announcing a new epoch.
However, if the process only moves \textit{full} blocks to the pool, then this limbo bag may be left with some records in its head block.
These records \textit{could} be moved to the pool immediately, but it is more efficient to leave them in the bag, and simply move them to the pool later, once the block that contains them is full.
One consequence of not moving these records to the pool is that each limbo bag can contain at most $B-1$ records that were retired two or more epochs ago.
This does not affect correctness.
The shared bag is implemented as a lock-free singly-linked list of blocks with operations to add and remove a full block. 
Moving entire blocks to and from the shared bag greatly reduces
synchronization costs. 

Operating on blocks instead of individual records significantly reduces overhead. 
However, it also requires a process to allocate and deallocate blocks.
To reduce the number of blocks that are allocated and deallocated during an execution, each process has a bounded \textit{block pool} that is used by all of its local blockbags.
Instead of deallocating a block, a process returns the block to its block bool.
If the block pool is already full, then the block is freed.
Experiments show that allowing each process to keep up to 16 blocks in its block pool reduces the number of blocks allocated by more than 99.9\%.
No blocks are allocated for the shared bag, since blocks are simply moved between pool bags and the shared bag.

\paragraph{Minor optimizations}

We make two additional optimizations in our implementation.
First, the least significant bit of \textit{announce}$_p$ is used as $p$'s quiescent bit.
This allows both values to be read and written atomically, which reduces the number of reads and writes to shared memory.
Second, a process attempts to increment the current epoch only after invoking \leaveq\ at least INCR\_THRESH times, where INCR\_THRESH is a constant (100 in our experiments).
This is especially helpful when the number of processes is small.
For example, in a single process system, without this optimization, the process will advance the epoch and try to move records to the pool at the beginning of every single operation, introducing unnecessary overhead.

\paragraph{Optimizing for NUMA systems}


Memory layout and access pattern can have a significant impact on the performance of an algorithm on a NUMA system.
%
Recall that each invocation of \leaveq\ by a process $q$ reads one announcement \textit{announce}$_p$ (which is periodically changed by process $p$).
If $p$ is on the same socket as $q$ (so $p$ and $q$ share the last-level cache), then this read will usually be quite fast, since a write by $p$ to \textit{announce}$_p$ will not invalidate $q$'s cached copy of \textit{announce}$_p$.
However, if they are on different sockets, then a write by $p$ to \textit{announce}$_p$ will invalidate $q$'s cached copy, and will cause a subsequent read of \textit{announce}$_p$ by $q$ to incur a last-level cache miss.
The cost of these last-level cache misses is not noticeable in all workloads, but we \textit{did} notice it in experimental workloads where all processes performed short read-only operations on small data structures that fit in the last-level cache (so the only last-level cache misses were caused by these reads).
To reduce the impact of these cache misses, we suggest reading an announcement only once every CHECK\_THRESH invocations of \leaveq\ (where CHECK\_THRESH is a small constant).
Of course, checking announcements less frequently can delay reclamation for some records.

\section{Adding fault tolerance} \label{sec-debraplus}

The primary disadvantage of DEBRA is that a crashed or slow process can stay in a non-quiescent state for an arbitrarily long time.
This prevents any other processes from freeing memory.
Although we did not observe this pathological behaviour in our experiments, many applications require a bound on the number of records waiting to be freed.


Lock-free data structures are ideal for building fault tolerant systems, because they are designed to be provably fault tolerant.
If a process crashes while it is in the middle of any lock-free operation, and it leaves the data structure in an inconsistent state, other processes can always repair that state.
The onus is on a process that wants to access part of a data structure to restore that part of the data structure to a consistent state before using it.
Consequently, a lock-free data structure always provides procedures to repair and access parts of the data structure that are damaged (by a process crash) or undergoing changes.
(Furthermore, these procedures are necessarily designed so that processes can crash while executing them, and other processes can still repair the data structure and continue to make progress.)
We use these procedures to design a mechanism that allows a process to \textit{neutralize} another process that is preventing it from advancing the epoch.

One novel aspect of DEBRA+ lies in its use of an inter-process communication mechanism called \textit{signaling}, which is offered by Unix, Linux and other POSIX-compliant operating systems.
Signals can be sent to a process by the operating system, and by other processes.
When a process receives a signal, the code it was executing is interrupted, and the process begins executing a \textit{signal handler}, instead.
When the process returns from the signal handler, it resumes executing from where it was interrupted.
A process can specify what action it will take when it receives a particular signal by registering a function as its signal handler. 

DEBRA+ also uses a feature of the C/C++ language called \textit{non-local goto}, which allows a process to begin executing from a different instruction, \textit{outside} of the current function, by using two procedures: \textit{sigsetjmp} and \textit{siglongjmp}.
A process first invokes \textit{sigsetjmp}, which saves its local state and returns false.
Later, the process can invoke \textit{siglongjmp}, which restores the state saved by \textit{sigsetjmp} immediately prior to its return, but causes it to return true instead of false.
The standard way to use these primitives is with the idiom: 
``if (sigsetjmp(...)) alternate(); else usual();''. 
A process that executes this code will save its state and execute usual().
Then, if the process later invokes \textit{siglongjmp}, it will restore the state saved by \textit{sigsetjmp} and immediately begin executing alternate().
Non-local goto is often used in the context of software transactional memory to implement transaction aborts.
Here, we combine non-local goto with signaling to implement \textit{neutralizing}.

At the beginning of each operation by a process $q$, $q$ invokes \textit{sigsetjmp}, following the idiom described above, and then proceeds with its operation.
Another process $p$ can interrupt $q$ by sending a signal to $q$. 
We design $q$'s signal handler so that, if $q$ was interrupted while it was in a quiescent state, then $q$ will simply return from the signal handler and resume its regular execution from wherever it was interrupted.
However, \textit{if $q$ was interrupted in a non-quiescent state, then it is neutralized}: it will enter a quiescent state and perform a \textit{siglongjmp}.
Then, $q$ will execute special \textit{recovery code}, which allows it to clean up any mess it left because it was neutralized.
Since $q$'s signal handler performs \textit{siglongjmp} only if $q$ was interrupted in a non-quiescent state, $q$ will not perform \textit{siglongjmp} while it is executing recovery code.
Hence, if $q$ receives a signal while it is executing recovery code, it will simply return from the signal handler and resume executing wherever it left off.
More generally, since $q$ will not be neutralized while it is quiescent, it can safely perform system calls, or any other code for which it does not have recovery code.


Our technique requires the operating system to guarantee that, after a process $p$ sends $q$ a signal, 
the next time process $q$ takes a step, it will execute its signal handler.
(This requirement is satisfied by the Linux kernel~\cite{kerrisk2010linux}. 
It can also be weakened with small modifications to DEBRA+, which are discussed at the end of this section.)
With this guarantee, after $p$ has sent a signal to $q$, it knows that $q$ will not access any retired record until $q$ has executed its recovery code and subsequently executed \leaveq.
Thus, as soon as $p$ has sent $q$ a signal, $p$ can immediately proceed as if $q$ is quiescent.

\paragraph{Operations for which recovery is simple}

\begin{wrapfigure}{r}{0.4\textwidth}
\vspace{-5mm}
\includegraphics[width=\linewidth]{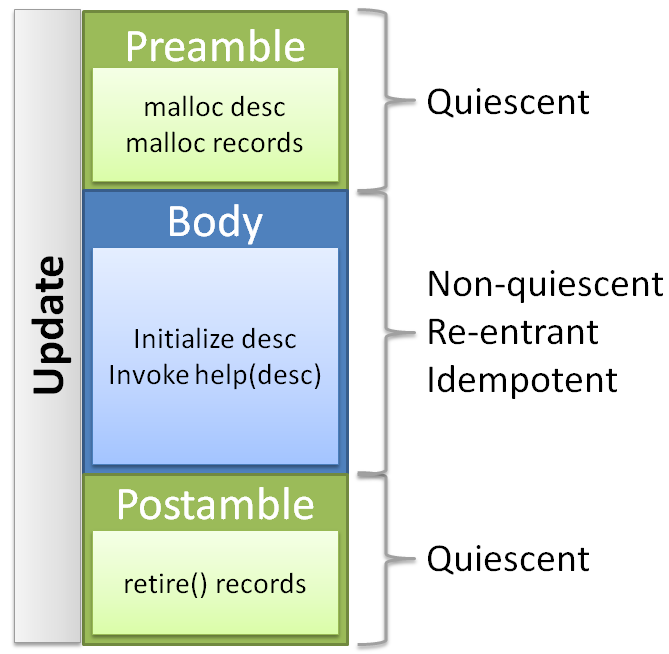}
\vspace{-12mm}
\end{wrapfigure}
The main difficulty in using DEBRA+ is designing recovery code.
Although recovery must be tailored to the data structure, it is straightforward for lock-free operations of the following form.
Each operation is divided into three parts: a quiescent bookkeeping \textit{preamble}, a non-quiescent \textit{body}, and a quiescent bookkeeping \textit{postamble}.
Processes can be neutralized while executing in the body, but cannot be neutralized while executing in the preamble or postamble (because a process will not call \textit{siglongjmp} while it is quiescent).
Consequently, processes should not do anything in the body that will corrupt the data structure if they are neutralized part way through. 
Allocation, deallocation, manipulation of process-local data structures that persist between operations, and other non-reentrant actions should occur only in the preamble and postamble.

\paragraph{Applying DEBRA+}

Figure~\ref{fig-applying-debraplus} shows how to apply DEBRA+ to the type of operations described above.
The remainder of this section explains the steps shown there.
Consider an operation $O$.
In the quiescent preamble, $O$ allocates a special record called a \textit{descriptor}. 
A pointer to this descriptor is stored in a process local variable called $desc$.
Any other records that might be needed by the body are also allocated in the preamble.

\begin{figure}
\begin{minipage}{\linewidth}
%
%
%
%
%
%
%
%
%
%
%
%
%
%
%
%
%
%
%
%
%
%
%
%
%
%
%
%
%
%
%
%
%
\lstset{escapechar=@,style=customc}
\begin{lstlisting}[name=applying,aboveskip=0pt,belowskip=0pt]
process local variables:
   descriptor *desc; @\vspace{1.5mm}@
\end{lstlisting}
\begin{lstlisting}[name=applying,aboveskip=0pt,belowskip=0pt,backgroundcolor=\color{gray!20}]
void signalHandler(args):
  if (isQuiescent()) then
    enterQstate();
    siglongjmp(...);        // jump to recovery code@\vspace{1.5mm}@
\end{lstlisting}
\begin{lstlisting}[name=applying,aboveskip=0pt,belowskip=0pt]
int doOperationXYZ(args):
  ...                       // quiescent preamble
  while (!done)
\end{lstlisting}
\begin{lstlisting}[name=applying,aboveskip=0pt,belowskip=0pt,backgroundcolor=\color{gray!20}]
    if (sigsetjmp(...))     // begin recovery code
      if (isRProtected(desc)) done = help(desc);
      RUnprotectAll();
    else                    // end recovery code
\end{lstlisting}
\begin{lstlisting}[name=applying,aboveskip=0pt,belowskip=0pt,backgroundcolor=\color{gray!0}]
      leaveQstate();        // begin body
\end{lstlisting}
\begin{lstlisting}[name=applying,aboveskip=0pt,belowskip=0pt]
      do search phase
      initialize *desc
\end{lstlisting}
\begin{lstlisting}[name=applying,aboveskip=0pt,belowskip=0pt,backgroundcolor=\color{gray!20}]
      RProtect each record that will be accessed, or used as the old value of a CAS, by help(desc)
      RProtect(desc);
\end{lstlisting}
\begin{lstlisting}[name=applying,aboveskip=0pt,belowskip=0pt]
      done = help(desc);
\end{lstlisting}
\begin{lstlisting}[name=applying,aboveskip=0pt,belowskip=0pt,backgroundcolor=\color{gray!0}]
      enterQstate();        // end body
\end{lstlisting}
\begin{lstlisting}[name=applying,aboveskip=0pt,belowskip=0pt,backgroundcolor=\color{gray!20}]
      RUnprotectAll();
\end{lstlisting}
\begin{lstlisting}[name=applying,aboveskip=0pt,belowskip=0pt]
  ...                       // quiescent postamble
  perform retire() calls
\end{lstlisting}
\end{minipage}
\vspace{-2mm}
\caption{
	Applying DEBRA+ to a typical lock-free operation.
	Lines 14 and 20 are necessary for both DEBRA and DEBRA+.
	Gray lines are necessary for fault tolerance (DEBRA+).
} 
\label{fig-applying-debraplus}
\end{figure}

The body first reads some records, and then initializes the descriptor $desc$.
Intuitively, this descriptor contains a description of all the steps the operation $O$ will perform.
The body then executes a \textit{help} procedure, which uses the information in the descriptor to perform the operation.
We assume that the descriptor includes all pointers that the \textit{help} procedure will follow (or use as the expected value of a CAS). 
The \textit{help} procedure can also be used by other processes to help the operation complete.
In a system where processes may crash, a process whose progress is blocked by another operation cannot simply wait for the operation to complete, so helping is necessary.
The \textit{help} procedure for any lock-free algorithm is typically reentrant and idempotent, because, at any time, one process can pause, and another process can begin helping.
The end of the body is marked with an invocation of \enterq\ (which is, itself, reentrant and idempotent).
The quiescent postamble then invokes \textit{retire} for each record that was removed from the data structure by the operation.


\paragraph{Recovery} 
We now describe recovery for an operation $O$ performed by a process $p$.
Suppose $p$ receives a signal, enters a quiescent state, and then performs \textit{siglongjmp} to begin executing its recovery code.
Although there are many places where $p$ might have been executing when it was neutralized, it is fairly simply to determine what action it should take next.
The main challenge is determining whether another process already performed $O$ on $p$'s behalf.
To do this, $p$ checks whether it announced a descriptor for $O$ before it was neutralized.
If it did, then some other process might have seen this descriptor and started helping $O$.
So, $p$ invokes \textit{help} (which is safe even if another process already helped $O$, since \textit{help} is idempotent).
Otherwise, 
$p$ can simply restart the body of $O$.


DEBRA allows a non-quiescent process executing an operation to safely access any record that it reached by following pointers from an entry point during the operation.
However, DEBRA does \textit{not} allow quiescent processes to safely access any records. 
In DEBRA+, once a process $p$ has been sent a signal, other processes treat $p$ as if were quiescent.
Furthermore, $p$ enters a quiescent state before executing recovery code, and it remains in a quiescent state throughout the recovery code.
However, the help procedure in $p$'s recovery code must access the descriptor record, and possibly some of the records to which it points. 
Thus, we need an additional mechanism in DEBRA+ to allow $p$ to access this limited set of \record s even though it is quiescent.
We use HPs in a very limited way to prevent these records from being freed by other processes before $p$ has finished 
its recovery code.
This lets $p$ safely run its recovery code in a quiescent state, so that other processes can continue to advance the current epoch and reclaim memory. 

We now describe how HPs are used.
Let $S$ be the set of records that will be accessed, or used as the old value of a CAS, by \textit{help}$(desc)$.
In the body of an operation by $p$, 
$p$ announces HPs to all records in $S$ by invoking \textit{RProtect}$(r)$ for each $r \in S$. 
Then, $p$ invokes \textit{RProtect}$(desc)$ to announce a HP to the descriptor, and invokes \textit{help}$(desc)$.
After performing \textit{help}$(desc)$ and invoking \enterq, $p$ invokes \textit{RUnprotectAll} to release all of its HPs.
Note that, since \textit{RProtect} is performed in the body, while $p$ is non-quiescent, $p$ might be neutralized while executing \textit{RProtect}.
Hence, \textit{RProtect} must be reentrant and idempotent.

When executing recovery code, $p$ first invokes \textit{isRProtected}$(desc)$ (which returns true if some HP points to $desc$ and false otherwise) to determine whether it announced a HP to $desc$.
Suppose it did.
Since $p$ announces a HP to the descriptor $d$ \textit{after} announcing HPs to all records in $S$, when $p$ performs recovery, if it sees that it announced a HP to $d$, then it knows it already announced HPs to all records in $S$.
Thus, its HPs will prevent everything it will access during its recovery from being reclaimed until it has finished using them.
So, $p$ can safely execute \textit{help}$(desc)$.
Now, suppose $p$ did not announce a HP to $desc$.
Since $p$ announces a HP to $desc$ \textit{before} invoking \textit{help}$(desc)$, this means $p$ has not yet invoked \textit{help}$(desc)$, so no other process is aware of $p$'s operation.
Therefore, $p$ can simply terminate its recovery code and restart its operation.
At the end of the recovery code, $p$ invokes \textit{RUnprotectAll} to release all of its HPs.

Recall that, in DEBRA, whenever a process $p$ announces a new epoch, it can immediately reclaim all records in its oldest limbo bag and move them to its pool.
In DEBRA+, some of the records in $p$'s oldest limbo bag might be pointed to by HPs, so $p$ cannot simply move all of these to its pool.
Before $p$ can move a record $r$ to the pool, it must first verify that no HP points to it. 
We discuss how this can be done efficiently, below.

\begin{figure}[ph]
\lstset{escapechar=@,style=customc}
\begin{lstlisting}[frame=single]
process local variables:
    hashtable scanning;                 // hash table used to collect all RProtected records
shared variables:
    arraystack RProtected[n];           // array of RProtected record* for each process

bool isRProtected(record *r) { return RProtected[pid].contains(r); }
bool RProtect(record *r)     { RProtected[pid].add(r);             } // O(1) time
void RUnprotectAll()         { RProtected[pid].clear();            } // O(1) time
bool leaveQstate() {
    bool result = false;
    long readEpoch = epoch;
    if (!isEqual(readEpoch, announce[other])) { // our announcement differs from the current epoch
        checkNext = 0;                    // we are now scanning announcements for a new epoch
        rotateAndReclaim();
        result = true;                  // result indicates that we changed our announcement
    }
    // incrementally scan all announcements
    if (++opsSinceCheck >= CHECK_THRESH) {
        opsSinceCheck = 0;
        long other = checkNext % n;
        if (isEqual(readEpoch, announce[other]) || isQuiescent(other) || suspectNeutralized(other)) {
            long c = ++checkNext;
            if (c >= n && c >= INCR_THRESH) { // if we have scanned every announcement
                CAS(&epoch, readEpoch, readEpoch+1);
    }   }   }
    announce[pid] = readEpoch;          // announce new epoch with quiescent bit = false
    return result;
}
void rotateAndReclaim() {
    index = (current+1) % 3;      // compute index of oldest limbo bag
    currentBag = bags[index];     // reuse the oldest limbo bag as the new currentBag
    // if currentBag contains sufficiently many records to get amortized O(1) time per record
    if (currentBag->getSizeInBlocks() >= scanThreshold) {
        // hash all announcements
        scanning.clear();
        for (int other=0; other < n; ++other) {
            int sz = RProtected[other].size();
            for (int i=0; i<sz; ++i) {
                record *hp = RProtected[other].get(i);
                if (hp != NULL) {
                    scanning.insert(hp);
        }   }   }
        // if any records in currentBag are RProtected, swap them to the front
        blockbag_iterator it1 = currentBag->begin();
        blockbag_iterator it2 = currentBag->begin();
        while (it1 != currentBag->end()) {
            if (scanning.contains(*it1)) {  // record pointed to by it1 is RProtected
                swap(it1, it2);             // swap records pointed to by it1 and it2
                it2++;                      // advance iterator it2
            }
            it1++;                          // advance iterator it1
        }
        // now, every record after it2 can be freed, so we reclaim all full blocks after it2
        pool->moveFullBlocks(it2);          // O(1) time
}   }
bool suspectNeutralized(long other) {
    return (currentBag->getSizeInBlocks() >= SUSPECT_THRESHOLD_IN_BLOCKS)
        && (!pthread_kill(getPthreadID(other), SIGQUIT)); // successfully send signal to other
}
\end{lstlisting}
\vspace{-2mm}
\lstset{escapechar=@,style=customc}
\begin{lstlisting}[frame=single]
void signalhandler(int signum, siginfo_t *info, void *uctx) {
    // if the process is not in a quiescent state, it jumps to a different instruction
    // and cleans up after itself, instead of continuing its current operation.
    if (!isQuiescent()) {
        enterQstate();
        siglongjmp(...);
}   } // otherwise, the process simply continues its operation as if nothing had happened.
\end{lstlisting}
\caption{C++ style pseudocode for data and procedures \textbf{added to DEBRA} to obtain DEBRA+, where $n$ is the number of processes, INCR\_THRESH is the minimum number of times a process must invoke \leaveq\ before it can increment the epoch, and CHECK\_THRESH is the number of times a process must invoke \leaveq\ before it will check the epoch announcement of another process.}
\label{fig-debraplus}
\end{figure}

\paragraph{Complexity}

Thanks to our new \textit{neutralizing} mechanism, we can bound the number of \record s waiting to be freed.
Each time a process $p$ performing \leaveq\ encounters a process $q$ that is not quiescent and has not announced epoch $e$, $p$ checks whether the size of its own current limbo bag exceeds some constant $c$.
If so, $p$ neutralizes $q$.
After $p$'s current limbo bag contains at least $c$ elements, and $p$ performs $n$ more data structure operations, it will have performed \leaveq\ $n$ times, and each non-quiescent process will either have announced the current epoch or been neutralized by $p$.
Consequently, $p$ will advance the current epoch, and, the next time it performs \leaveq, it will announce the new epoch and reclaim records.
It follows that $p$'s current limbo bag can contain at most $c+O(nm)$ elements, where $m$ is the largest number of \record s that can be removed from the data structure by a high-level operation.
Therefore, the total number of \record s waiting to be freed is $O(n(nm+c))$.

In DEBRA, all full blocks in a limbo bag are moved to the pool in constant time.
In DEBRA+, \record s can be moved to the pool only if no HP points to them, so this is no longer possible.
One way to move \record s from a limbo bag $b$ to the pool is to iterate over each \record\ $r$ in $b$, and check if a HP points to $r$.
To make this more efficient, we move records from $b$ to the pool only when $b$ contains $nk+\Omega(nk)+B$ \record s, where $k$ is the number of HPs needed per process, and $B$ is the maximum number of \record s in a block.
Before we begin iterating over \record s in $b$, we create a hash table containing every HP.
Then, we can check whether a HP points to $r$ in O(1) expected time.
We can further optimize by rearranging \record s in $b$ so that we can still move full blocks to the pool, instead of individual records.
To do this, we iterate over the \record s in $b$, and move the ones pointed to by HPs to the beginning of the blockbag.
All full blocks that do not contain a \record\ pointed to by a HP are then moved in O(1) time.
Since there are at most $nk$ HPs, and we scan only when $b$ contains at least $nk+B+\Omega(nk)$ \record s, we will be able to move at least $max\{B, \Omega(nk)\}$ \record s to the pool.
Thus, the expected amortized cost to move a record to the pool (or free it) is O(1).

\trevor{Give EXAMPLE here of applying DEBRA+ to a simple d.s. that uses LLX/SCX/template?}

\paragraph{Implementation}

C++ style pseudocode for DEBRA+ appears in Figure~\ref{fig-debraplus}.
There, only the procedures that are different from DEBRA are shown.
There are three main differences from DEBRA.
First, in an invocation of \leaveq\ by process $p$, if $p$ encounters a process $q$ that has not announced the current epoch, and is not quiescent, $p$ invokes a procedure called \textit{suspectNeutralized}.
This procedure checks whether $p$'s current limbo bag contains more than a certain number of records, and, if so, neutralizes $q$.
Recall that $p$ neutralizes $q$ by sending a signal to $q$.
The signal handler that is executed by $q$ when it receives a signal is called \textit{signalhandler}.
Second, a limited version of HPs is provided by procedures \textit{isRProtected}, \textit{RProtect} and \textit{RUnprotectAll}.
Third, the procedure \textit{rotateAndReclaim} implements the algorithm described above efficiently moving records from a limbo bag to a pool (only if the records are not pointed to by HPs) in expected amortized O(1) steps per record.

One problem with DEBRA+ is that it seems difficult to apply to lock-based data structures, because it is dangerous to interrupt a process and force it to restart while it holds a lock.
Of course, there is little reason to use DEBRA+ for a lock-based data structure, since locks can cause deadlock if processes crash. 
For lock-based data structures, DEBRA can be used, instead.

\paragraph{Alternative implementation options}

Above, we specified
the guarantee that the operating system signaling mechanism must provide: after a process $p$ sends process $q$ a signal, the next time $q$ takes a step, it will execute its signal handler.
It is possible to modify DEBRA+ to work with a weaker guarantee.
Suppose the operating system instead guaranteed that, after a process $p$ sends process $q$ a signal, $q$ is guaranteed to begin executing its signal handler when it next experiences a context switch.
Then, after $p$ sends a signal to $q$, it can either wait or defer reclamation until $q$ is next context switched or is not running.
For most operating system schedulers, every process is context switched out after a bounded length of time (a scheduling quantum).
Many operating systems also provide mechanisms to determine whether a process is currently running, how many context switches it has undergone, how much time remains until it will be context switched, and so on.
For example, Linux provides access to this information through a virtual file system (rooted at "/proc").


%



\section{A lock-free memory management abstraction} \label{sec-abstraction}

There are many compelling reasons to separate memory allocation and reclamation from data structure code.
Although the steps that a programmer must take to apply a non-automatic technique to a data structure vary widely, it is possible to find a small, natural set of operations that allows 
a programmer to write data structure code once, and easily plug in many popular memory reclamation schemes.
In this section, we describe a record management abstraction, called a Record Manager, that is easy to use, and provides sufficient flexibility to support: HPs (all versions), B\&C, TS, EBR, DTA, QS, DEBRA and DEBRA+.
(ST is not supported because it requires a programmer to insert transactions throughout the code, and to annotate the beginning and end of each stack frame.
OA is not supported because it requires each read and CAS to be instrumented.)


\begin{wrapfigure}{r}{0.5\textwidth}
\vspace{-5mm}
\includegraphics[width=\linewidth]{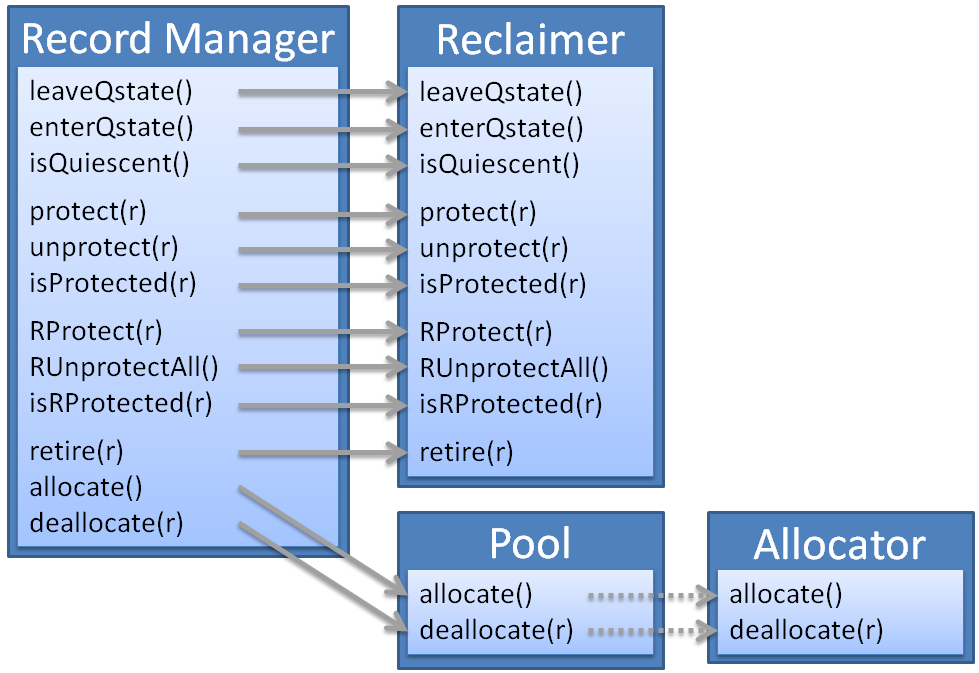}
\caption{Operations provided by a Record Manager, Reclaimer, Pool and Allocator. Solid arrows (resp., dashed arrows) indicate that an operation on one object invokes (resp., may invoke) an operation on another object.}
\label{fig-record-manager-operations}
\end{wrapfigure}
A Record Manager has three components: an Allocator, a Reclaimer and a Pool. 
The Allocator determines how records will be allocated (e.g., by individual calls to \textit{malloc} or by handing out records from a large range of memory) and freed.
The Reclaimer is given records after they are removed from the data structure, and determines when they can be safely handed off to the Pool.
The Pool determines when records are handed to the Allocator to be freed, and whether a process actually uses the Allocator to allocate a new record.

We implement data structures, Allocators, Reclaimers and Pools in a modular way, so that they can be combined easily.
This clean separation into interchangeable components allows, e.g., the same Pool implementation to be used with both a HP Reclaimer and a DEBRA Reclaimer.
Modularity is typically achieved with inheritance, but inheritance introduces significant runtime overhead.
For example, when the precompiled data structure invokes \textit{retire}, it does not know which of the precompiled versions of \textit{retire} it should use, so it must perform work at runtime to choose the correct implementation.
In C++, this can be done more efficiently with \textit{template parameters}, which allow a compiler to reach into 
code and replace placeholder calls with calls to the correct implementations.
Unlike inheritance, templates introduce no overhead, since the correct implementation is compiled into the code.
Furthermore, if the correct implementation is a small function, the compiler can simply insert its code directly into the calling function (eliminating the function call altogether).

A programmer interacts with the Record Manager, which exposes the operations of the Pool and Reclaimer (depicted in Figure~\ref{fig-record-manager-operations}).
A Pool provides \textit{allocate} and \textit{deallocate} operations.
A Reclaimer provides operations for the basic events that memory reclamation schemes are interested in: starting and finishing data structure operations (\leaveq\ and \enterq), reaching a new pointer and disposing of it (\textit{protect} and \textit{unprotect}), and retiring a record (\textit{retire}).
It also provides operations to check whether a process is quiescent (\isq) and whether a pointer can be followed safely (\textit{isProtected}).
Finally, it provides operations for making information available to recovery code (\textit{RProtect}, \textit{RUnprotectAll}, \textit{isRProtected}).
%

Most of these operations are described in Section~\ref{sec-technique} and Section~\ref{sec-debraplus}.
We describe the rest here.
\textit{protect}, which must be invoked on a record $r$ before accessing any field of $r$, returns true if the process successfully protects $r$ (and, hence, is permitted to access its fields), and returns false otherwise.
Once a process has successfully protected $r$, it remains protected until the process invokes \textit{unprotect}$(r)$ or becomes quiescent.
\textit{isProtected}$(r)$ returns true if $r$ is currently protected by the process.

Reclaimers for DEBRA and DEBRA+ are effectively described in Section~\ref{sec-technique} and Section~\ref{sec-debraplus}. 
For these techniques, \textit{unprotect} does nothing, and \textit{protect} and \textit{isProtected} simply return true.
(Consequently, these calls are optimized out of the code by the compiler.)
For HPs, \leaveq, \textit{RProtect} and \textit{RUnprotectAll} all do nothing, and \isq\ and \textit{isRProtected} simply return false.
\textit{unprotect}$(r)$ releases a HP to $r$, and \enterq\ clears all announced HPs.
\textit{protect} announces a HP to a record and executes a function, which determines whether that record is in the data structure.
\textit{retire}$(r)$ places $r$ in a bag, and, if the bag contains sufficiently many records, it constructs a hash table $T$ containing all HPs, and moves all records not in $T$ to the Pool (as described in Section~\ref{sec-debra-related}). 

A predicate called \textit{supportsCrashRecovery} 
is added to Reclaimers to allow a programmer to add crash recovery to a data structure without imposing overhead for Reclaimers that do not support crash recovery.
For example, a programmer can check whether \textit{supportsCrashRecovery} is true before invoking \textit{RProtect}.
The code statement ``if (\textit{supportsCrashRecovery})'' 
statically evaluates to ``if (true)'' or ``if (false)'' at compile time, once the Reclaimer template has been filled in. 
Consequently, these if-statements are completely eliminated by the compiler. 
In our experiments, this predicate is used to invoke \textit{sigsetjmp} 
only for DEBRA+ (eliminating overhead for the other techniques). 


\section{Experiments} \label{sec-debra-exp}

Our primary experimental system was an Intel i7 4770 machine with 4 cores, 8 hardware threads and 16GB of memory, running Ubuntu 14.04 LTS.
All code was compiled with GCC 4.9.1-3 and the highest optimization level (-O3).
Google's high performance Thread Caching malloc (tcmalloc-2.4) was used.

We ran experiments to compare the performance of various Reclaimers: DEBRA, DEBRA+, HP, ST and no reclamation (None).
We used the Record Manager abstraction to perform allocation and reclamation for a lock-free balanced binary search tree (BST)~\cite{Brown:2014}.
Searches in this BST can traverse pointers from retired nodes to other retired nodes, so ST cannot be used, and we must confront the problems described in Section~\ref{sec-debra-related} to apply HP.
Properly dealing with HP's problems would be highly complex and inefficient, so we simply restart any operation that suspects a node is retired.
Consequently, applying HP causes the BST to lose its lock-free progress guarantee.
To determine whether this significantly affects the performance of HP, we added the same restarting behaviour to DEBRA, and observed that its impact on performance was small.
Note that the HP scheme was tuned for high performance (instead of space efficiency) by allowing processes to accumulate large buffers of retired nodes before attempting to reclaim memory.

Code for ST was graciously provided by its authors.
They used a lock-based skip list to compare None, HP and ST.
We modified their code to use a Record Manager for allocating and pooling nodes, and used it 
to compare None, DEBRA, HP and ST.
The actual reclamation code for HP and ST is due to the authors of ST.
Since the skip list uses locks, it cannot use DEBRA+. 

\begin{figure}[t]
    \begin{minipage}{0.5\textwidth}
    \setlength\tabcolsep{0pt}
    \centering
    \begin{tabular}{m{0.04\linewidth}m{0.48\linewidth}m{0.48\linewidth}}
        &
        \multicolumn{2}{c}{\fcolorbox{black!80}{black!40}{\parbox{\dimexpr 0.96\linewidth-2\fboxsep-2\fboxrule}{\centering\textbf{Experiment 1 on 8-thread Intel i7-4770}}}}
        \\
        &
        \fcolorbox{black!50}{black!20}{\parbox{\dimexpr \linewidth-2\fboxsep-2\fboxrule}{\centering {\footnotesize 50\% ins, 50\% del}}} &
        \fcolorbox{black!50}{black!20}{\parbox{\dimexpr \linewidth-2\fboxsep-2\fboxrule}{\centering {\footnotesize 25\% ins, 25\% del, 50\% search}}}
        \\
        \rotatebox{90}{{\footnotesize \textbf{BST range} $[0, 10^6)$}} &
        \includegraphics[width=\linewidth]{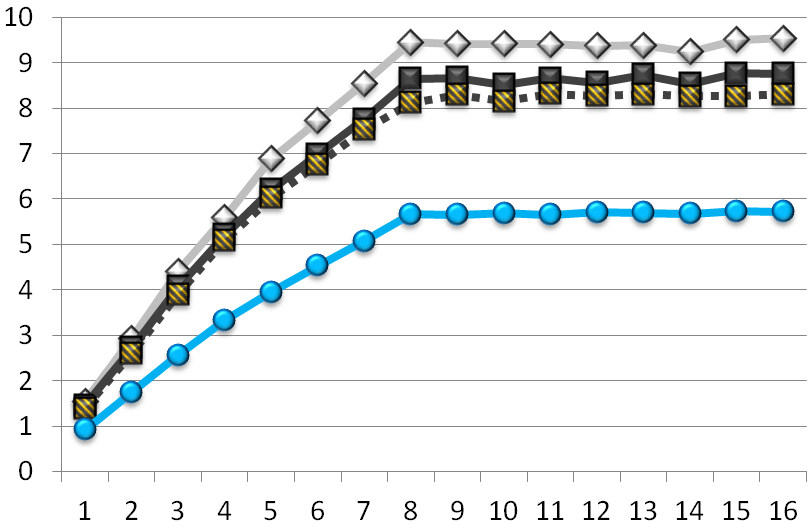} &
        \includegraphics[width=\linewidth]{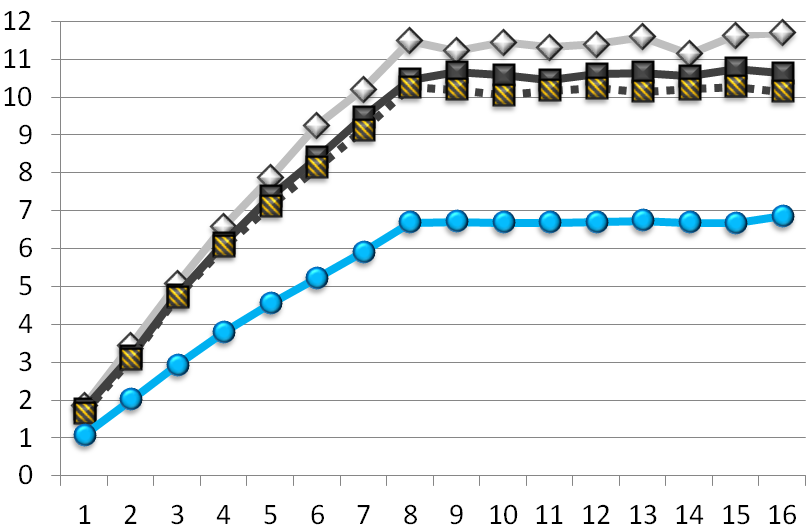}
        \\
        \rotatebox{90}{{\footnotesize \textbf{BST range} $[0, 10^4)$}} &
        \includegraphics[width=\linewidth]{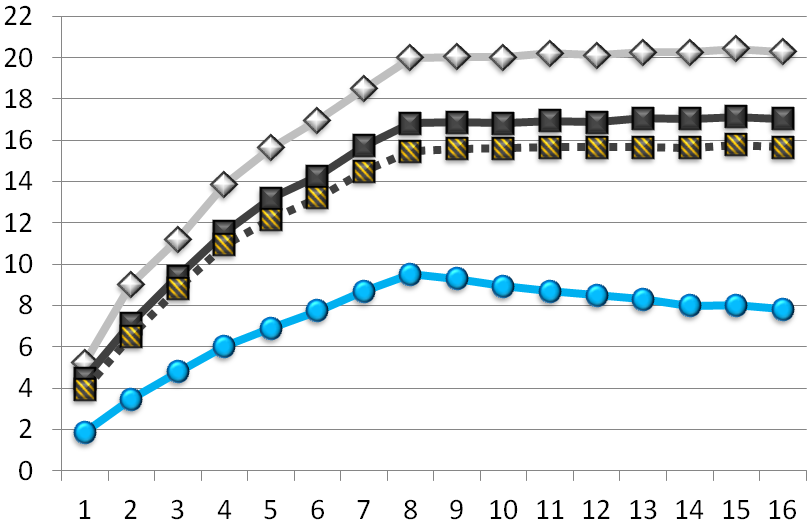} &
        \includegraphics[width=\linewidth]{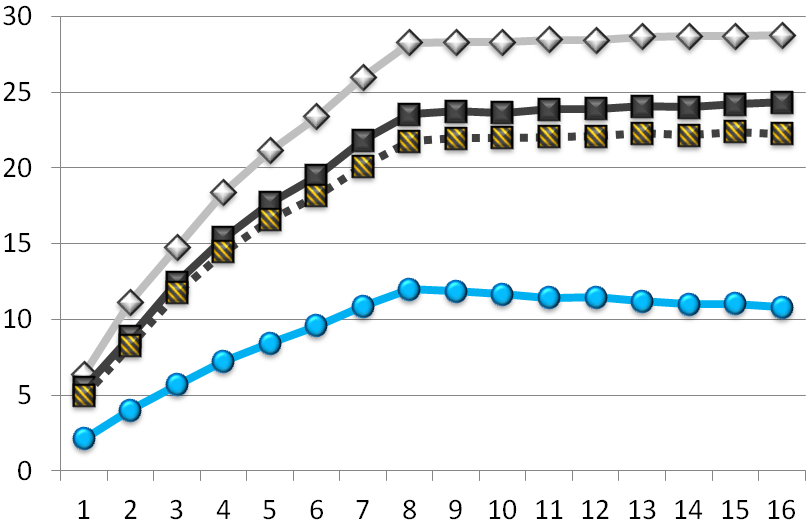}
        \\
        \rotatebox{90}{{\footnotesize \textbf{Skiplist range} $[0, 2 \cdot 10^5)$}} &
        \includegraphics[width=\linewidth]{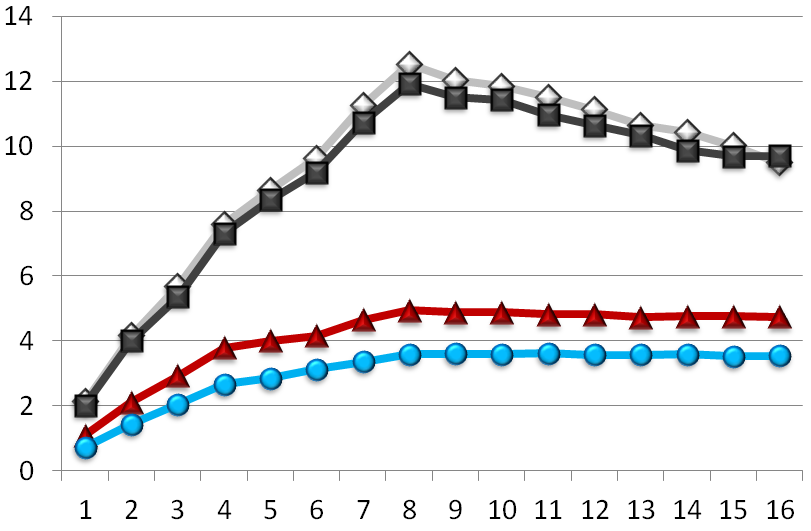} &
        \includegraphics[width=\linewidth]{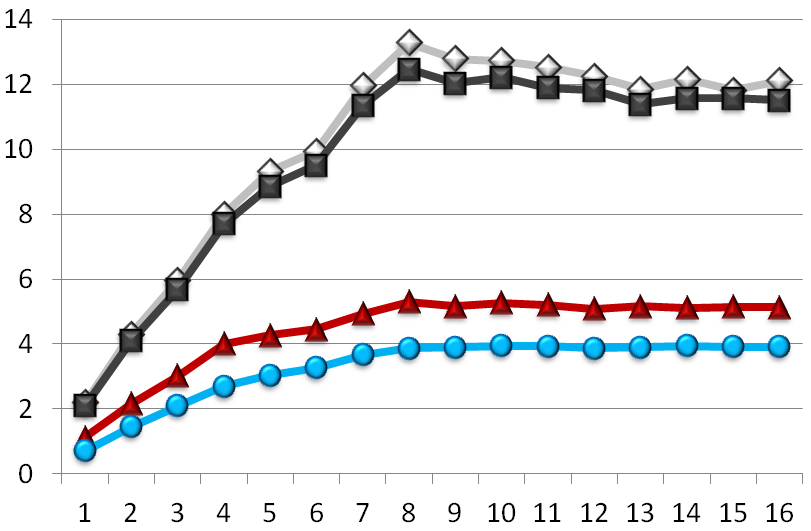}
        \\
        &
        \multicolumn{2}{c}{\includegraphics[width=0.96\textwidth]{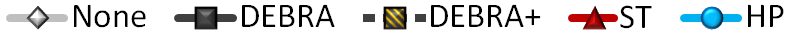}}
        \\
    \end{tabular}
    \end{minipage}
    \begin{minipage}{0.5\textwidth}
    \setlength\tabcolsep{0pt}
    \centering
    \begin{tabular}{m{0.04\linewidth}m{0.48\linewidth}m{0.48\linewidth}}
        &
        \multicolumn{2}{c}{\fcolorbox{black!80}{black!40}{\parbox{\dimexpr 0.96\linewidth-2\fboxsep-2\fboxrule}{\centering\textbf{Experiment 2 on 8-thread Intel i7-4770}}}}
        \\
        &
        \fcolorbox{black!50}{black!20}{\parbox{\dimexpr \linewidth-2\fboxsep-2\fboxrule}{\centering {\footnotesize 50\% ins, 50\% del}}} &
        \fcolorbox{black!50}{black!20}{\parbox{\dimexpr \linewidth-2\fboxsep-2\fboxrule}{\centering {\footnotesize 25\% ins, 25\% del, 50\% search}}}
        \\
        \rotatebox{90}{{\footnotesize \textbf{BST range} $[0, 10^6)$}} &
        \includegraphics[width=\linewidth]{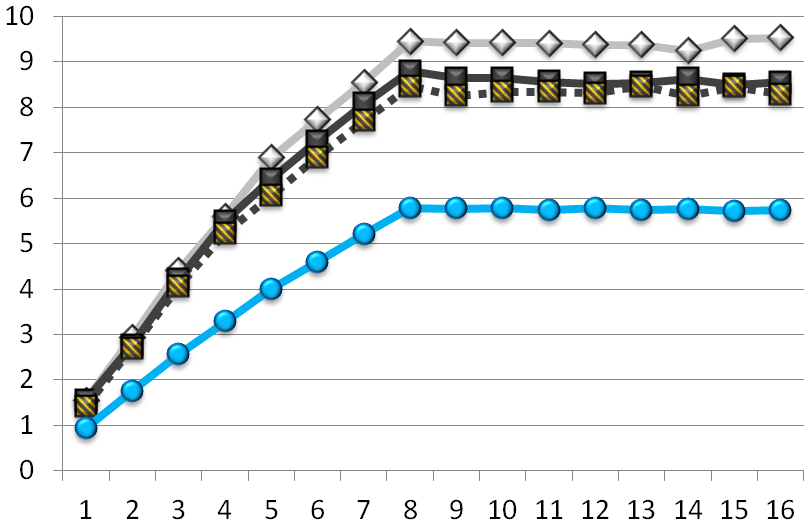} &
        \includegraphics[width=\linewidth]{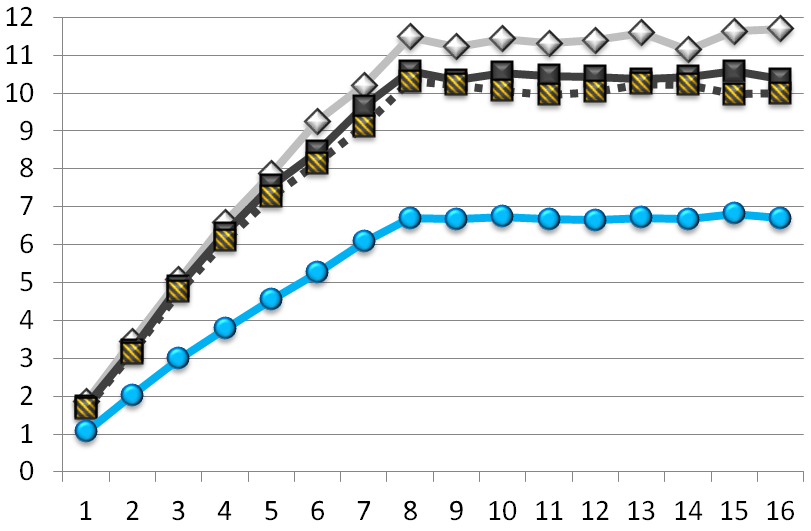}
        \\
        \rotatebox{90}{{\footnotesize \textbf{BST range} $[0, 10^4)$}} &
        \includegraphics[width=\linewidth]{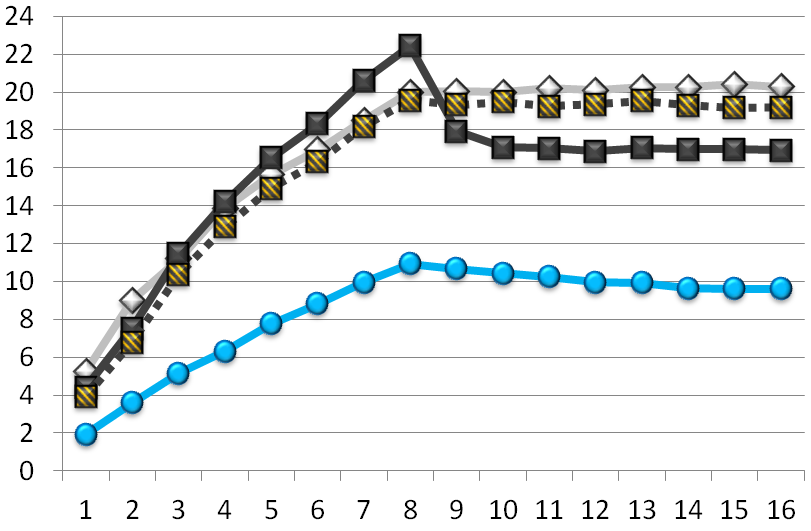} &
        \includegraphics[width=\linewidth]{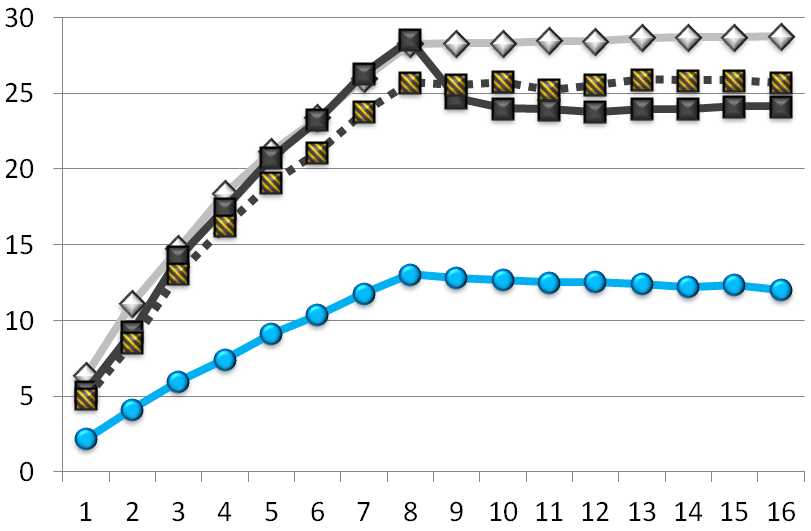}
        \\
        \rotatebox{90}{{\footnotesize \textbf{Skiplist range} $[0, 2 \cdot 10^5)$}} &
        \includegraphics[width=\linewidth]{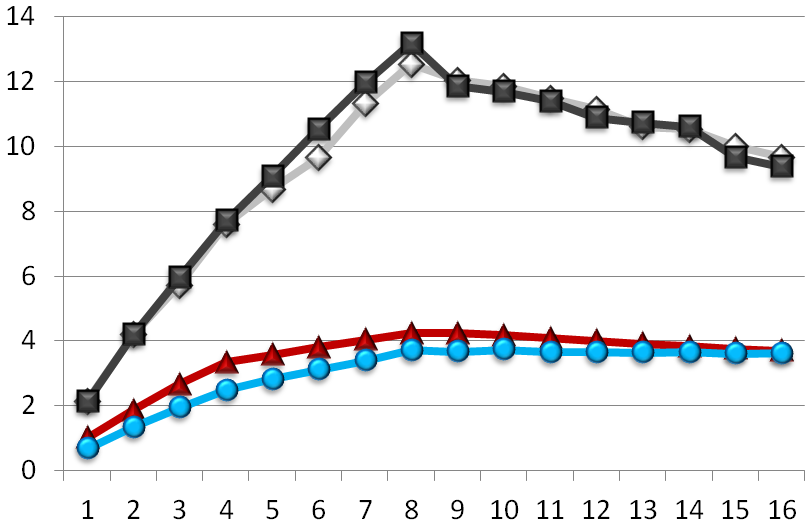} &
        \includegraphics[width=\linewidth]{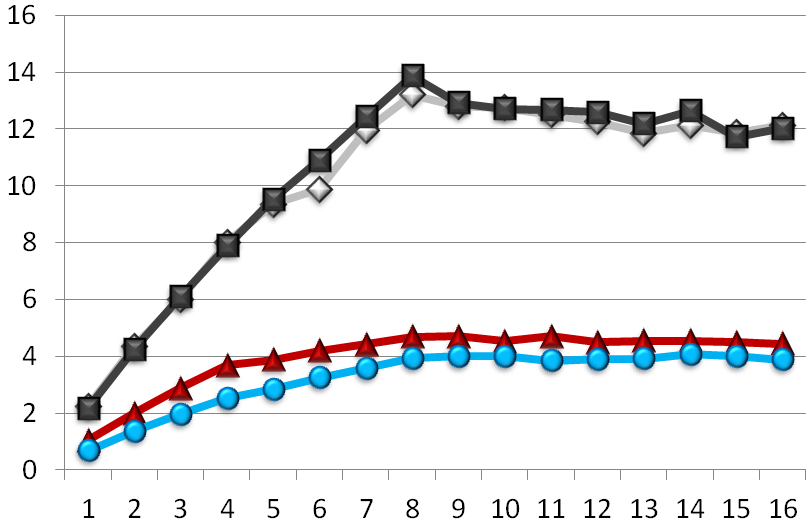}
        \\
        &
        \multicolumn{2}{c}{\includegraphics[width=0.96\textwidth]{chap-debra/graphs/legend.png}}
        \\
    \end{tabular}
    \end{minipage}
    \vspace{-2mm}
	\caption{
		Left: Experiment 1 (Overhead of reclamation).
        Right: Experiment 2 (Using a Bump Allocator and a Pool).
		The x-axis shows the number of processes.
		The y-axis shows throughput, in millions of operations per second.
	}
	\label{fig-exp1}
\end{figure}

\paragraph{Experiment 1}
Our first experiment compared the overhead of performing reclamation for the various Reclaimers.
In this experiment, each Reclaimer performed all the work necessary to reclaim nodes, but nodes were not actually reclaimed (and, hence, were not reused).
The Record Manager used a \textit{Bump Allocator}: each process requests a large region of memory from the operating system at the beginning of an execution, and then divides that region into nodes, which it allocates in sequence.
Since nodes were not actually reclaimed, we eliminated the Pool component of the Record Manager.
In this experiment, a data structure suffers the overhead of reclamation, but does \textit{not} enjoy its benefits (namely, a smaller memory footprint and fewer cache misses).

For the balanced BST, we ran eight \textit{trials} for each combination of Reclaimers in \{None, DEBRA, DEBRA+, HP\}, thread counts (in \{1, 2, ..., 16\}), operation mixes in \{25i-25d, 50i-50d\} (where $x$i-$y$d means $x$\% insertions, $y$\% deletions and ($100-x-y$)\% searches) and key ranges in \{[0, 10000), [0, 1000000)\}.
For the skip list, the thread counts and operation mixes were the same, but ST was used instead of DEBRA+, and there was only one key range, [0, 200000).
In each trial, the data structure was first prefilled to half of the key range, then the appropriate number of threads performed random operations (according to the operation mix) on uniformly random keys from the key range for two seconds.
The average of each set of eight trials became a data point in a graph.
(Unfortunately, the system quickly runs out of memory when nodes are not reclaimed, so it is not possible to run all trials for longer than two seconds.
However, we ran long trials for many cases to verify that the results do not change.)

The results in Figure~\ref{fig-exp1}(left) show that DEBRA and DEBRA+ have extremely low overhead.
In the BST, DEBRA has between 5\% and 22\% overhead (averaging 12\%), and DEBRA+ has between 7\% and 28\% overhead (averaging 17\%).
Compared to HP, on average, DEBRA performs 94\% more operations and DEBRA+ performs 83\% more.
This is largely because DEBRA and DEBRA+ synchronize once \textit{per operation}, whereas HP synchronizes each time a process reaches a new node.
In the skip list, DEBRA has up to 6\% overhead (averaging 4\%), outperforms HP by an average of 200\%, and also outperforms ST by between 93\% and 168\% (averaging 133\%).
ST has significant overhead.
For instance, on average, it starts almost four transactions per operation (each of which announces one or more pointers), and runtime checks are frequently performed to determine if a new transaction should be started.

\paragraph{Experiment 2}

In our second experiment, nodes were actually reclaimed.
The Reclaimers were each paired with the same Pool as DEBRA.
The only exception was None, which does not use a Pool.
(Thus, None is the same as in the first experiment.) 

\begin{figure}[t]
    \begin{minipage}{0.55\textwidth}
    \setlength\tabcolsep{0pt}
    \centering
    \begin{tabular}{m{0.04\linewidth}m{0.48\linewidth}m{0.48\linewidth}}
        &
        \multicolumn{2}{c}{\fcolorbox{black!80}{black!40}{\parbox{\dimexpr 0.96\linewidth-2\fboxsep-2\fboxrule}{\centering\textbf{Experiment 2 on 64-thread Oracle T4-1}}}}
        \\
        &
        \fcolorbox{black!50}{black!20}{\parbox{\dimexpr \linewidth-2\fboxsep-2\fboxrule}{\centering {\footnotesize 50\% ins, 50\% del}}} &
        \fcolorbox{black!50}{black!20}{\parbox{\dimexpr \linewidth-2\fboxsep-2\fboxrule}{\centering {\footnotesize 25\% ins, 25\% del, 50\% search}}}
        \\
        \rotatebox{90}{{\footnotesize \textbf{BST range} $[0, 10^6)$}} &
        \includegraphics[width=\linewidth]{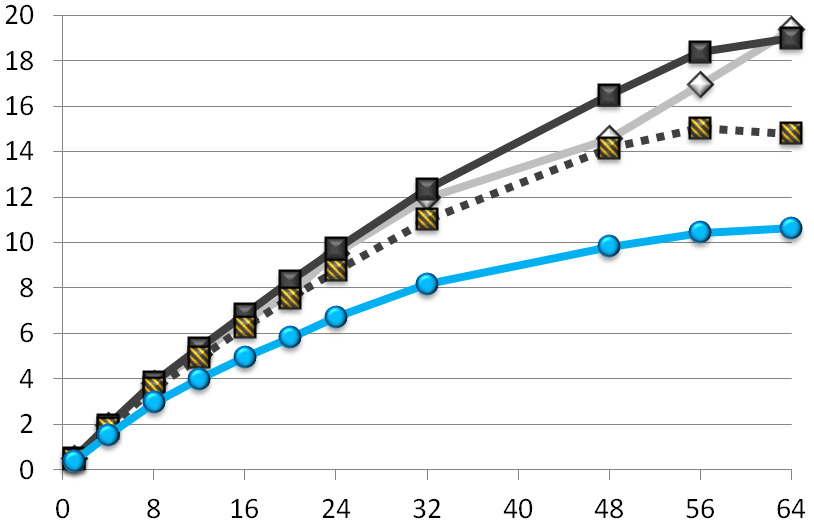} &
        \includegraphics[width=\linewidth]{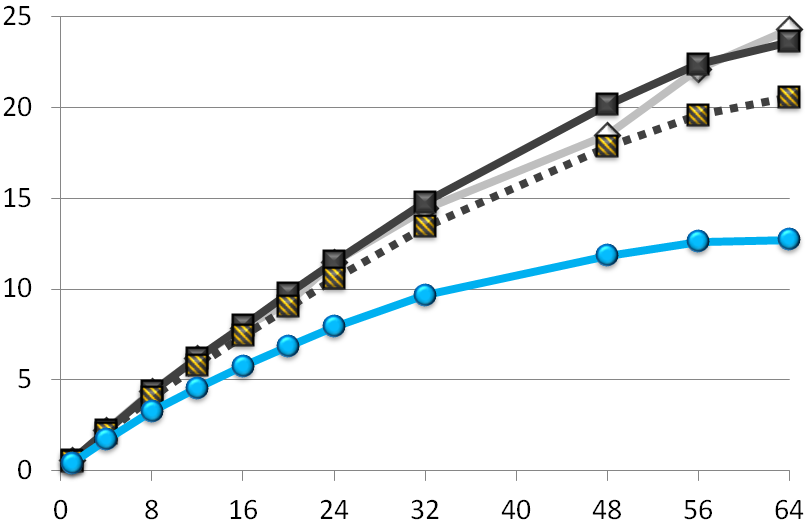}
        \\
        &
        \multicolumn{2}{c}{\includegraphics[width=0.96\textwidth]{chap-debra/graphs/legend.png}}
        \\
    \end{tabular}
    \end{minipage}
    \begin{minipage}{0.44\textwidth}
    \includegraphics[width=\linewidth]{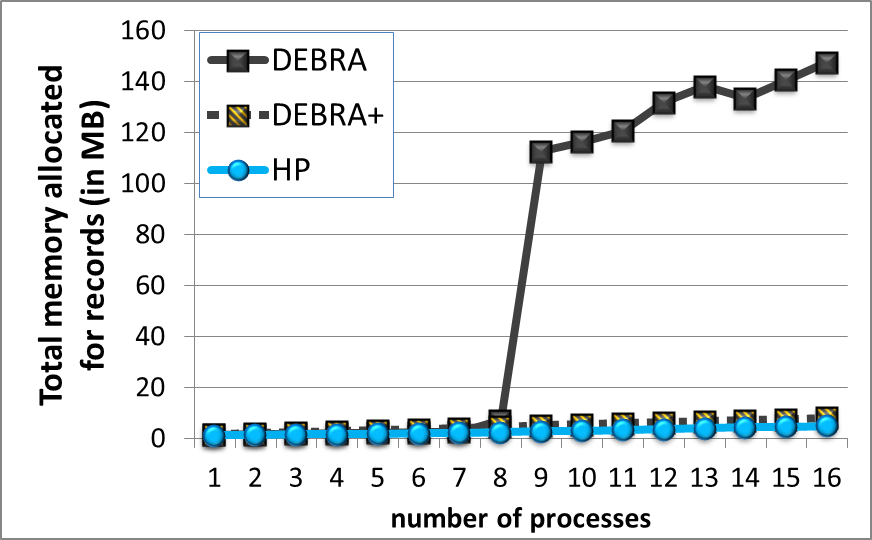}
    \end{minipage}
    \vspace{-2mm}
	\caption{
		(Left) Extra results for Experiment 2 on Oracle T4-1.
        (Right) Memory allocated for records in Experiment~2 in the BST with keyrange 10,000 and workload 50i-50d.
	}
	\label{fig-oracle-exp2}
\end{figure}


The results appear in Figure~\ref{fig-exp1}(right).
In the BST, DEBRA is only 8\% slower than None on average, and, for some data points, DEBRA actually improves performance by up to 12\%.
This is possible because DEBRA reduces the memory footprint of the data structure, which allows a larger fraction of the allocated nodes to fit in cache and, hence, reduces the number of cache misses.
DEBRA+ is between 2\% and 25\% slower than None, averaging 10\%.
Compared to HP, DEBRA is between 48\% and 145\% faster, averaging 80\%, and DEBRA+ is between 43\% and 123\% faster, averaging 76\%.
In the skip list, DEBRA performs \textit{as well as None}.
DEBRA also \textit{outperforms} ST by between 108\% and 211\%, averaging 160\%.


To measure the benefit of neutralizing slow processes, we tracked the total amount of memory allocated for records in each trial.
Since we used bump allocation, this simply required determining how far each bump allocator's pointer had moved during the execution.
Thus, we were able to compute the total amount of memory allocated \textit{after} each trial had finished (without having any impact on the trial while it was executing).
Figure~\ref{fig-oracle-exp2}(right) shows the total amount of memory allocated for records in the second experiment in the BST with key range 10,000 and workload 50i-50d.
(The other cases were similar.)
DEBRA, DEBRA+ and HP all perform similarly up to eight processes.
\begin{wrapfigure}{r}{0.5\textwidth}
    \centering
    \begin{minipage}{\linewidth}
    \setlength\tabcolsep{0pt}
    \centering
    \begin{tabular}{m{0.04\linewidth}m{0.48\linewidth}m{0.48\linewidth}}
        &
        \multicolumn{2}{c}{\fcolorbox{black!80}{black!40}{\parbox{\dimexpr 0.96\linewidth-2\fboxsep-2\fboxrule}{\centering\textbf{Experiment 3 on 8-thread Intel i7-4770}}}}
        \\
        &
        \fcolorbox{black!50}{black!20}{\parbox{\dimexpr \linewidth-2\fboxsep-2\fboxrule}{\centering {\footnotesize 50\% ins, 50\% del}}} &
        \fcolorbox{black!50}{black!20}{\parbox{\dimexpr \linewidth-2\fboxsep-2\fboxrule}{\centering {\footnotesize 25\% ins, 25\% del, 50\% search}}}
        \\
        \rotatebox{90}{{\footnotesize \textbf{BST range} $[0, 10^6)$}} &
        \includegraphics[width=\linewidth]{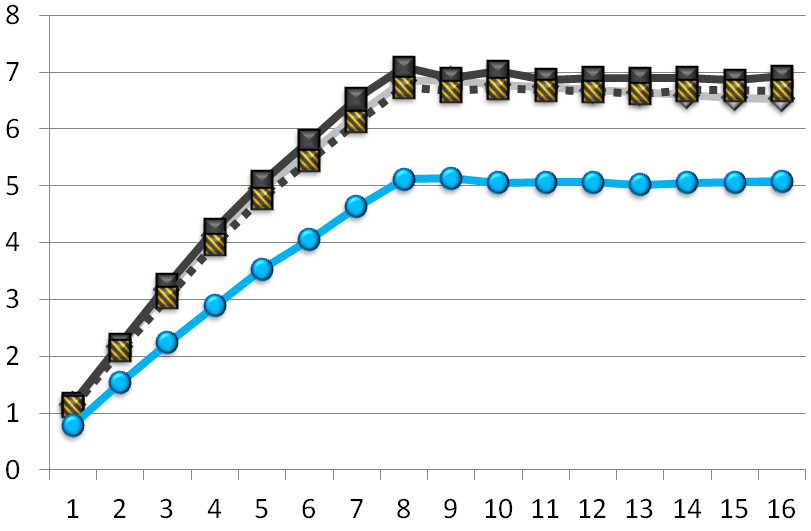} &
        \includegraphics[width=\linewidth]{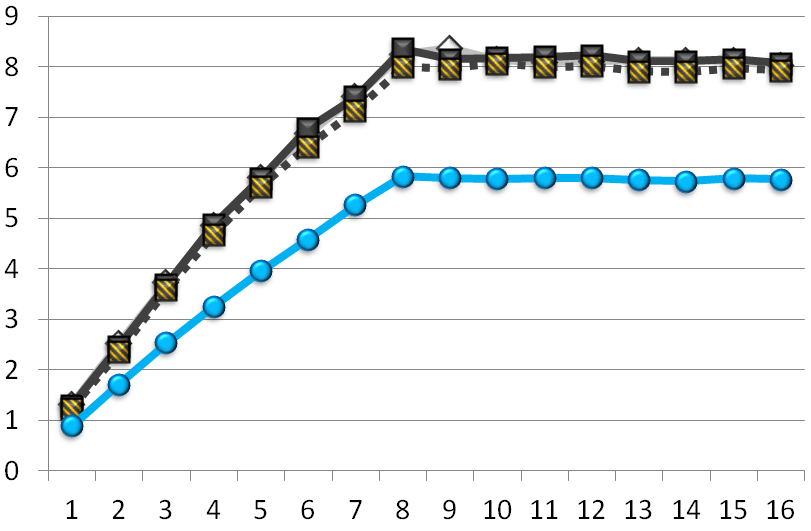}
        \\
        \rotatebox{90}{{\footnotesize \textbf{BST range} $[0, 10^4)$}} &
        \includegraphics[width=\linewidth]{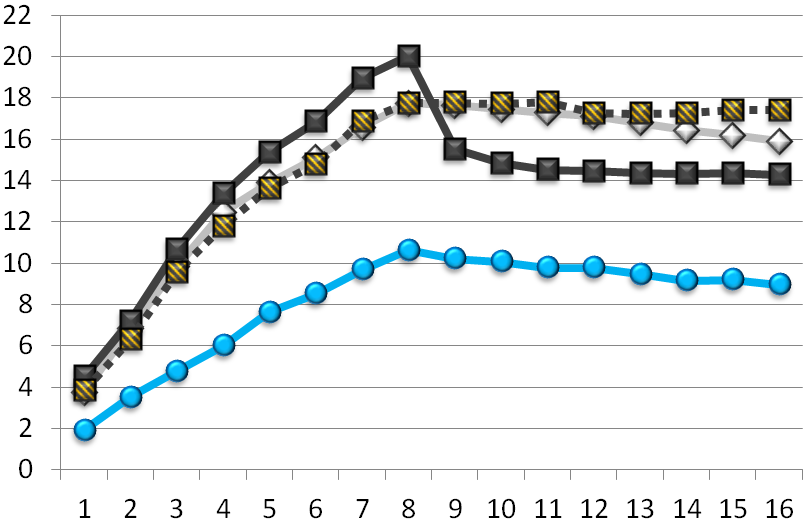} &
        \includegraphics[width=\linewidth]{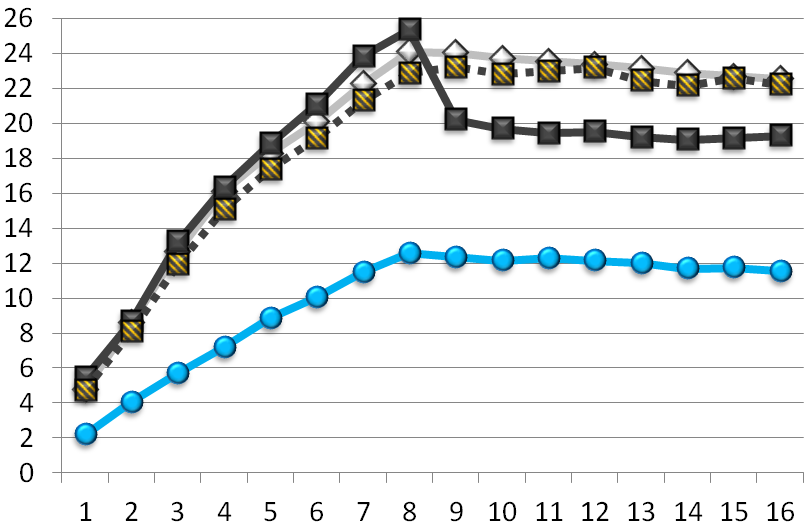}
        \\
        \rotatebox{90}{{\footnotesize \textbf{Skiplist range} $[0, 2 \cdot 10^5)$}} &
        \includegraphics[width=\linewidth]{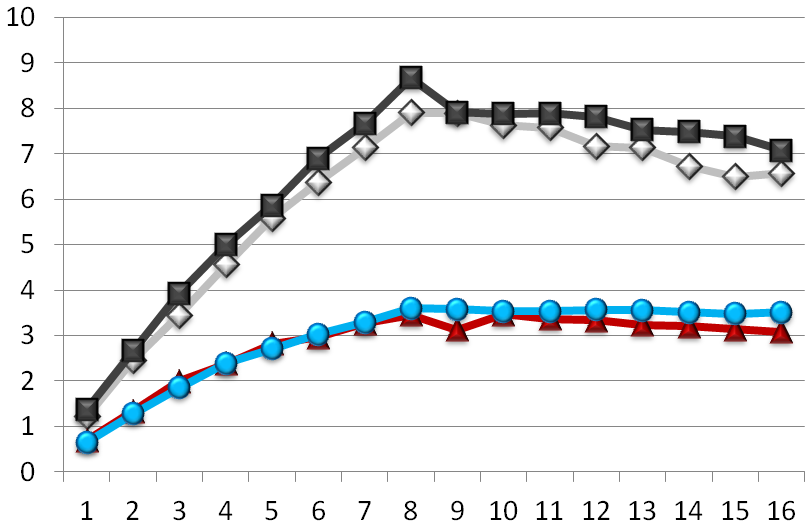} &
        \includegraphics[width=\linewidth]{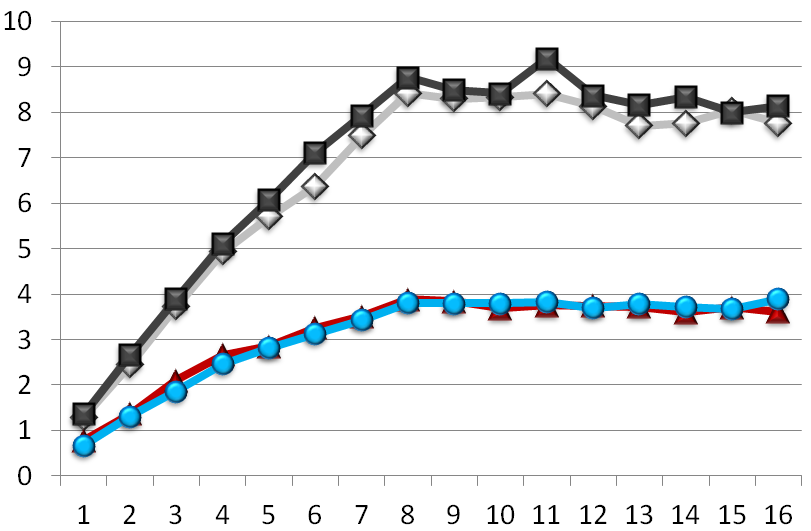}
        \\
        &
        \multicolumn{2}{c}{\includegraphics[width=0.96\textwidth]{chap-debra/graphs/legend.png}}
        \\
    \end{tabular}
    \end{minipage}
    \vspace{-3mm}
	\caption{
		Experiment 3 (Using malloc and a Pool).
        The x-axis shows the number of processes.
		The y-axis shows throughput, in millions of operations per second.
	}
	\label{fig-exp3}
\end{wrapfigure}
However, for more than eight processes, some processes are always context switched out, and they often prevent DEBRA from advancing the epoch in a timely manner.
DEBRA+ fixes this issue.
With 16 processes, DEBRA+ neutralizes processes an average of 935 times per trial, reducing peak memory usage by an average of 94\% over DEBRA.

We also ran the second experiment on a large scale NUMA Oracle T4-1 system with 8 cores and 64 hardware contexts.
Figure~\ref{fig-oracle-exp2}(left) shows a representative sample of the results.
Note that ST could not be run on this machine, since it does not support HTM.

\paragraph{Experiment 3}
Our third experiment is like the second, except we used a different Allocator, which does not preallocate memory.
The Allocator's \textit{allocate} operation simply invokes \textit{malloc} to request memory from the operation system (and its \textit{deallocate} operation invokes \textit{free}).


The results (which appear in Figure~\ref{fig-exp3}) are similar to the results for the second experiment.
However, the absolute throughput is significantly smaller than in the previous experiments, because of the overhead of invoking \textit{malloc}.
Although HP and ST are negatively affected, proportionally, they slow down less than None, DEBRA and DEBRA+.
This illustrates an important experimental principle: \textit{overhead should be minimized, because uniformly adding overhead to an experiment disproportionately impacts low-overhead algorithms, and obscures their advantage}.

\section{Conclusion}

In this work, we presented a distributed variant of EBR, called DEBRA.
Compared to EBR, DEBRA significantly reduces synchronization overhead and offers high performance even with many more processes than physical cores. 
Our experiments show that, compared with performing no reclamation at all, DEBRA is 4\% slower on average, 21\% slower at worst, and up to 20\% \textit{faster} in some cases.
Moreover, DEBRA outperforms StackTrack by an average of 138\%.
DEBRA is easy to use, and only adds O(1) steps per data structure operation and O(1) steps per retired record. 

We also presented DEBRA+, the first epoch based reclamation scheme that allows processes to continue reclaiming memory after a process has crashed.
In an $n$ process system, 
the number of objects waiting to be freed is $O(mn^2)$, where $m$ is the largest number of objects retired by one data structure operation.
The cost to reuse or free a record is O(1) expected amortized time. 
In our experiments, DEBRA+ reduced memory consumption over DEBRA by 94\%.
Compared with performing no reclamation, DEBRA+ is only 10\% slower on average.
DEBRA+ also outperforms a highly efficient implementation of hazard pointers by an average of 70\%. 

We introduced the \textit{Record Manager}, the first generalization of the C++ \textit{Allocator} abstraction that is suitable for lock-free programming.
A Record Manager separates memory reclamation code from lock-free data structure code, which allows a dynamic data structure to be implemented without knowing how its records will be allocated, reclaimed and freed.
This abstraction adds virtually no overhead.
It is highly flexible, allowing a programmer to interchange techniques for reclamation, object pooling, allocation and deallocation by changing one line of code.
%
Complete C++ code for DEBRA and DEBRA+, the Record Manager abstraction, and all data structure implementations studied in this work, is available at \url{http://implementations.tbrown.pro}.

Besides DEBRA and DEBRA+, the neutralizing technique introduced in this work is of independent interest.
It would be useful to find different ways to neutralize processes, so, for example, the neutralizing technique could be used with different operating systems.
There may also be opportunities to apply neutralizing in other contexts, such as garbage collection. 
Finally, it would also be interesting to understand whether these ideas can be extended to lock-based algorithms (even for a restricted class, such as reentrant and idempotent algorithms).

\subsection*{Acknowledgments}

This work was performed while the author was a student at the University of Toronto.
Funding was provided by the Natural Sciences and Engineering Research Council of Canada.
The author would also like thank his supervisor Faith Ellen for her helpful comments on this work, to Oracle Labs for providing the 64-thread SPARC machine, and to Dave Dice for running the experiments on that machine.

\bibliographystyle{abbrv}
\bibliography{bibliography}

\end{document}